\documentclass[useAMS, usegraphicx]{mn2e}
\usepackage{amssymb}

\newcommand{\ebv}{$E(B-V)$}

\newcommand{\Msun}{ $M_{\odot}$}

\newcommand{\z}{$z$}

\newcommand{\zs}{$z$$\sim$}

\newcommand{\zphot}{$z_{phot}$}
\newcommand{\sqam}{arcmin$^2$}

\newcommand{\OmOlHo}{$(\Omega_M, \Omega_\Lambda, H_0)$}
\newcommand{\cosmoparams}{(0.3, 0.7, 70 km s$^{-1}$ Mpc$^{-1}$)}

\newcommand{\R}{$\cal R$}

\newcommand{\Rlim}{${\cal R}_{lim}$}

\newcommand{\Uhdf}{$U_{300}$}
\newcommand{\Bhdf}{$B_{450}$}
\newcommand{\Vhdf}{$V_{606}$}
\newcommand{\Ihdf}{$I_{814}$}
\newcommand{\Rave}{${\cal R}_{VI}$}
\newcommand{\Jhdf}{$J_{110}$}
\newcommand{\Hhdf}{$H_{160}$}
\newcommand{\hdffilters}{$U_{300}B_{450}V_{606}I_{814}J_{110}H_{160}$}

\newcommand{\s}{$\sim$}
\newcommand{\Mstar}{$M^*$}
\newcommand{\Lstar}{$L^*$}

\newcommand{\phistar}{$\phi^*$}
\newcommand{\chisq}{$\chi^2$}
\newcommand{\Muv}{$M_{1700}$}
\newcommand{\Veff}{$V_{eff}$}

\newcommand{\kdfdata}{KDF~I}
\newcommand{\kdflf}{KDF~II}
\newcommand{\kdfldens}{KDF~III}

\newcommand{\sfe}{$f_*$}
\newcommand{\lcdm}{$\Lambda$CDM}
\newcommand{\Mstars}{$M_{stars}$}

\title[UV-Selected Sub-L* Galaxies at \zs2]{Stars, Dust, and the Growth of UV-Selected Sub-L* Galaxies at Redshift \zs2}
\author
[Marcin Sawicki]{Marcin Sawicki$^{1}$\thanks{E-mail: sawicki@ap.smu.ca} 
\\
$^{1}$Department of Astronomy and Physics, Saint Mary's University, 923 Robie Street, Halifax, Nova Scotia, B3H 3C3, Canada\\
}
\begin{document}

\date{Accepted . Received ; in original form}

\pagerange{\pageref{firstpage}--\pageref{lastpage}} \pubyear{2011}

\maketitle

\label{firstpage}

\begin{abstract}
This work concerns the physical properties of very faint ($R_{lim}$=28 AB mag; $M_{stars, lim}$$\sim$$10^8$\Msun), UV-selected sub-\Lstar\ BX galaxies at \zs2.3. Stellar masses, dust content, and dust-corrected star formation rates are constrained using broadband spectral energy distribution fitting, resulting in a number of insights into the nature of these low-mass systems. First, a correlation between rest-frame UV luminosity and galaxy stellar mass appears to exist in BX galaxies and its presence suggests that many sub-\Lstar\ galaxies at this redshift may have approximately constant, rather than highly variable, star formation histories. A nearly-linear relation between stellar mass and star formation rate is also found, hinting that the rate at which a sub-\Lstar\ BX galaxy forms its stars is directly related to the mass of stars that it has already formed. A possible explanation for this phenomenon lies in a scenario in which new gas that falls onto the galaxy's host halo along with accreting dark matter is the main source of fuel for ongoing star formation. The instantaneous efficiency of star formation is low in this scenario, of order one percent. Turning to bulk quantities, it is found that the low-mass end of the stellar mass function at \zs2.3 is steeper than expected from extrapolations of shallower surveys, resulting in a stellar mass density at \zs2.3 that's  $\sim$25\% of the present-day value; this value is $\sim$1.5--2 times higher than that given by extrapolations of most of the shallower surveys, suggesting that the build-up of stellar mass in the universe has proceeded somewhat more rapidly than previously thought. With SED-fitting results in hand, an update to the Keck Deep Fields \zs2 UV luminosity function finds a steeper faint-end slope than previously reported, $\alpha=-1.47$, though this is not as steep as that found by Reddy \& Steidel (2009).  Finally, it is also found that sub-\Lstar\ galaxies at \zs2 carry very small amounts of dust compared to their more luminous cousins, so that while only $\sim$20\% of 1700\AA photons escape from a typical \Mstar\ galaxy, more than half make it out of an \Mstar+3 one. This paucity of dust highlights the fact that sub-\Lstar\ galaxies are not simple scaled copies of their more luminous cousins. Assuming that absorption by neutral hydrogen is not stronger in sub-\Lstar\ galaxies than in their more luminous counterparts, it also means that sub-\Lstar\ are important contributors to keeping the Universe ionized at \zs2.

\end{abstract}


\section{Introduction}

The epoch around \zs2 is particularly important for our understanding of galaxy formation and evolution: it is at this time, when the Universe was only about a quarter of its present age, that the rates of star formation and AGN activity were at their peak, and when many key processes, such as downsizing and the formation of the red sequence of massive galaxies,  were likely getting underway.   However, to date, studies of \zs2 galaxies have focused largely on luminous members of the population. This focus has in part been motivated by theoretical interests in massive systems as a means to test hierarchical galaxy formation models, but part of the problem is also practical: because high-$z$ galaxies are faint and thus challenging to observe, most follow-up studies to date have focused on relatively luminous objects around the L* ÒkneeÓ in the luminosity function (LF). While economical, this approach has limited us to studying what are, in fact, quite rare objects. 

Only recently has the importance of sub-\Lstar\ galaxies started to be appreciated. It is now becoming apparent that sub-\Lstar\ galaxies at high redshift are important contributors to the star formation and metal production budgets in the cosmos and contain a significant fraction of its assembled stellar mass (e.g., Sawicki \& Thompson 2006; Reddy \& Steidel 2009).  Given the paucity of quasars and luminous galaxies, sub-\Lstar\ galaxies are also postulated as the dominant sources of cosmic reionization at very high redshifts, z$>$5 (e.g., Bouwens et al. 2009). Meanwhile, it is unwise to assume that sub-\Lstar\ galaxies are simple scaled-down copies of their super-\Lstar\ cousins --- instead, they may differ from them in key properties such star formation histories, metallicities,  dust content, and so forth.  Indeed, the shape of the galaxy luminosity function, and the presence of phenomena such as downsizing  (e.g., Cowie et al. 1996) point to different evolutionary mechanisms in sub-\Lstar\ and super-\Lstar\ populations --- differences that are also expected from theoretical considerations of mechanisms for regulating and suppressing star formation (e.g., supernova vs. stellar feedback) as function of dark halo mass. It is thus vital that we understand the nature of sub-\Lstar\ galaxies at high redshift, how they differ from the better studied $\sim$\Lstar\ objects, and what they tell us about how galaxies form and evolve. 

Several techniques for selecting galaxies at \zs2 are now available (e.g., Franx et al. 2003; Daddi et al.\ 2004). Galaxies selected using the LBG/BM/BX selection techniques (Steidel et al.\ 1999, 2003, 2004), which are based on rest-frame UV broadband photometry and thus sensitive to star-forming populations that are at most only weakly obscured by interstellar dust, are among the best studied and understood.  In order to leverage on this knowledge, the present work pushes the BX selection to much fainter objects than those that are usually studied. 

As is the case with more luminous galaxies, once a sample of galaxies is in hand, several lines of attack are available.  The luminosity function (e.g., Gabasch 2004; Sawicki \& Thompson 2006; Reddy \& Steidel 2009) provides a useful but crude descriptor of the galaxy population, and although it is important to measure the LF's faint end accurately, the shape of the LF and its evolution has only limited interpretive value.  Spectroscopic observations would be desirable, but for sub-\Lstar\ galaxies at \zs2 they are extremely challenging in terms of telescope time and consequently little is know about the spectral properties of these important objects. Galaxy-galaxy clustering can tell us about the dark matter halos that host high-redshift galaxies (e.g., Adelberger et al.\ 1998, 2005; Ouchi et al.\ 2001), but such work is only now being done for sub-\Lstar\ galaxies at \zs2 (Savoy, Sawicki, Thompson, \& Sato 2011).  Finally, much can be learned from broadband spectral energy distributions of high-$z$ galaxies (e.g., Sawicki \& Yee 1998, Papovich et al.\ 2001, Shapley et al.\ 2001); this paper presents the results of such a study of the stellar populations of UV-selected faint (\Rlim = 28 AB) galaxies at \zs2. 

This paper is organized as follows.  Section~\ref{sec:data} describes the source data, photometry, and sample selection of the sub-L* \zs2 sample. Section~\ref{sec:SEDfitting} deals with estimation of stellar masses, dust-corrected star formation rates (SFRs), and other parameters using Spectral Energy Distribution (SED) fitting.  Results of the SED fitting are presented in \S~\ref{sec:results}. Section~\ref{sec:discussion} contains a discussion of the broader implications of the results on our understanding of galaxy formation and evolution.  Finally, \S~\ref{sec:conclusions} summarizes the key results. Throughout this work it is assumed that \OmOlHo\ = \cosmoparams.  At $z$=2.3, this cosmology gives the age of the Universe to be 2.8 Gyr. Unless otherwise noted, all magnitudes are on the AB system (Oke 1974). 

\section{Data}\label{sec:data}

\begin{figure*}
\begin{center}
  \includegraphics[height=0.350\textheight]{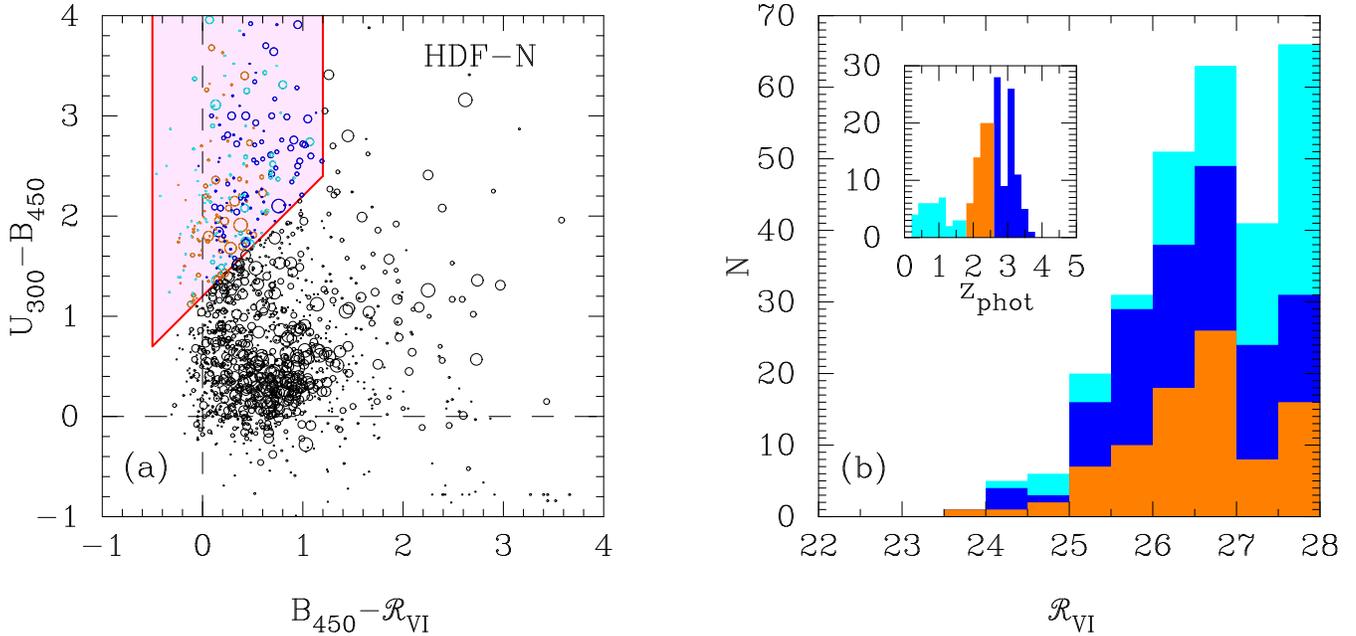}
\end{center}
 \caption[]{Sample selection.  The {\it left panel} shows the color-color diagram for all the HDF objects with 23$<$\Rave$<$28. Larger symbols correspond to brighter objects. The color-color selection region (Eq.~\ref{eq:color-color} and Steidel et al.\ 1996) is highlighted and the objects within this region are color-coded according to their photometric redshifts, as given in the right-hand panel. The photometric redshift distribution is shown in the inset plot in the {\it right panel}, with the three colors corresponding to the three ranges defined by Eq.~\ref{eq:zphotcut}.  The right panel also shows the number counts of the color-color-selected objects: the lowest, orange histogram shows the numbers for the final sample of 91 objects, blue denotes the objects with \zphot$>$2.6, and cyan those with \zphot$<$1.8.  Note that the histograms are stacked on top of each other, rather than overlapping.}
\label{sampleselection.fig}
\end{figure*}

\subsection{The Hubble Deep Field: Oldie but Goodie}

This study uses the publically available WFPC2 and NICMOS data from the original Hubble Deep Field (HDF; Williams et al.\ 1996; Dickinson 1999).  While several excellent multi-wavelength HST datasets, such as the GOODS fields or the Hubble Ultra Deep Field (HUDF), surpass the original HDF in both depth and/or area, the depth of \Uhdf\ imaging in the original HDF maintains its position as an excellent dataset for studying star-forming galaxies at $z$$\sim$2.  This is because selection of UV-bright galaxies at \zs2 can be done efficiently using a combination of observed-frame UV/optical filters that span $\sim$3000-6000\AA, with deep $U$-band imaging being critical. (e.g., Steidel et al.\ 1996, 2004; Adelberger et al.\ 2004).  As Steidel et al.\ (2004) have shown, color-color selection that includes the $U$ band identifies \zs2 galaxies while rejecting most foreground ($z$$\la$1.5) interlopers.  The very deep \Uhdf\ data of the HDF allow such \zs2 selection, making it excellent for studying sub-L* galaxies at this epoch. 

The HDF has an area of only $\sim$5.3~\sqam, but this relatively small size is not a serious limitation in the present work.  While studies of the luminosity function and of the clustering properties of galaxies need large samples drawn from large volumes as they need to combat simple Poisson noise in the counts as well as clustering-induced field-to-field variance ("cosmic variance"), these are not the dominant sources of noise in SED studies. Instead, what is critically important in SED-fitting work are data with very deep and well-calibrated photometry, while the number of galaxies in the sample need only be large enough to allow the identification of major trends in their physical parameters.  With several dozen \zs2 galaxies (see \S~\ref{sec:the_sample}) and very deep multicolor data, from \Uhdf\ through to \Hhdf, the HDF is well-suited for such work. 

\subsection{Object detection and photometry}

Object detection and photometry were done using SExtractor version 2.3.2 (Bertin \& Arnouts 1996) on the publically available WFPC (\Uhdf \Bhdf \Vhdf \Ihdf) and NICMOS (\Jhdf \Hhdf) image mosaics of the HDF.  The \Vhdf\ image is the deepest among the six bands, and so object detection was carried out on the unsmoothed \Vhdf-band images, requiring for detection that an object have a minimum of 10 pixels brighter than the 3$\sigma$ sky level. The HDF images are extremely deep, and faint high-$z$ UV-selected galaxies are compact (e.g., Giavalisco, Steidel \& Macchetto 1996), so it is possible to reach faint levels in the population.  Simulations that use compact but not point-like sources show that the present catalog is $\sim$50\% complete at \Vhdf = 29.1 and $>$90\% complete at \Vhdf=28.8.

Accurate colors are critical for SED studies and great care was taken to measure them reliably.  First, the unsmoothed \Vhdf\ image was used to measure the SExtractor total \Vhdf-band magnitudes.  This step needs only be accurate to $\sim$10\% and so SExtractor MAG\_AUTO magnitudes were sufficient here.   In contrast, the measurement of flux ratios (i.e., colors) between bands for a given object are much more critical for estimating accurate stellar masses, ages, dust corrections, and dust-corrected star formation rates.  To ensure accurate color measurements, photometry in the other bands was done on smoothed \hdffilters\ images as follows.  First, the \hdffilters\ images were smoothed using Gaussian kernels to give common FWHM of $\sim$0.17\arcsec\ that matches the FWHM in the lowest-resolution image, namely the \Hhdf.  Next, SExtractor was run in its dual-image mode, in which it detected objects in the \emph{unsmoothed} \Vhdf\ image, but carried out photometry at the corresponding location in the smoothed \hdffilters\ images. Here, fluxes were measured using positionally-matched, fixed circular apertures of 12 pixels ($\sim$0.5\arcsec) diameter.  These aperture magnitudes were then transformed into ``total'' magnitudes by applying the offsets found between the \Vhdf-band total and aperture magnitudes: $m_{tot} = m_{ap} + (V_{606, tot} - V_{606, ap})$, where the $m_i$ refer to the total and aperture magnitudes in the various bands \Uhdf \Bhdf \Ihdf \Jhdf \Hhdf.  Using images smoothed to a common resolution ensures that the 12-pixel ``color apertures" capture the same fraction of a galaxy's flux in each waveband, and results in accurate galaxy colors that are essential for SED fitting.  Measurements on compact objects in the smoothed data show that colors measured within the 12-pixel apertures in the smoothed are typically accurate to $\sim$1\% or better, without incurring large sky noise that would result from using larger apertures.  Overall, the above procedure results in a catalog of \Vhdf-selected objects with ``total" magnitudes that accurately represent object colors.

The Steidel et al.\ (1996, 2004) selection of \zs2 galaxies relies on $R$-band data, and so it was necessary to construct a synthetic $R$-band magnitude for each object.   Following Steidel et al.\ (1996), this was done by averaging the \Vhdf\ and \Ihdf\ fluxes: 
\begin{equation}
\label{defRave.eq}
{\cal R}_{VI} = -2.5 \log \Big(\frac{10^{-0.4V_{606}} + 10^{-0.4I_{814}}}{2}\Big).
\end{equation}
The sample retained for further analysis was then defined on the basis of this synthetic \Rave\ magnitude and consists of 1498 objects with \Rave$<$28.0.  At this relatively bright (for the HDF) magnitude cutoff the sample is close to complete and photometric errors for the majority of objects are still only a few percent in most bands.

\subsection{The $z$$\sim$2.3 sample}\label{sec:the_sample}

The \zs2.3 sample is magnitude limited and is selected using color-color LBG selection followed by a photometric redshift cut, as defined below.  These steps mimic the selection  procedures used to select $z$$\sim$2.3 ``BX'' galaxy samples (e.g., Steidel et al.\ 2004; Shapley et al.\ 2005; Erb et al.\ 2006; Reddy et al.\ 2007).  The main differences between the Steidel group's samples and the one used here are the slightly different filter systems, and the use of photometric redshifts --- instead of spectroscopic ones --- in the present, much fainter sample.  These differences are small and the present HDF sample selects a population very similar to the Steidel et al.\  \zs2.3 ``BX'' galaxies, while pushing to much fainter depths, with a catalog limit at \R=28 instead of the \R=25.5 as typically used by the Steidel et al.\ team for SED fitting work (Shapley et al.\ 2005; Erb et al.\ 2006). 

The sample is based on HDF HST data, and consequently it seemed reasonable to use the color-color selection criteria defined by Steidel et al.\ (1996) for their Hubble Deep Field high-$z$ galaxy selection.  These criteria are: 
\begin{eqnarray}
\label{eq:color-color} U_{300} - B_{450} > 1.2 + (B_{450} - {\cal R}_{VI}), \\ 
-0.5 < (B_{450} - {\cal R}_{VI}) < 1.2. \nonumber 
\end{eqnarray}
In addition to the color-color criteria, the sample is restricted to the magnitude cut  defined by 
\begin{equation}
\label{eq:magcut}
23<{\cal R}_{VI}<28.
\end{equation}
This magnitude cut is design to guard against potential rare, anomalous objects at the bright end, and, at the faint end, to exclude objects with large photometric errors that propagate to large errors on the SED fitting output parameters. Note that the sample is nearly complete at these relatively bright magnitudes.

The left panel of Fig.~\ref{sampleselection.fig} shows the colors of all the 23 $\leq$ \Rave\ $\leq$ 28 HDF objects in the \Uhdf \Bhdf \Rave\ color-color plane, together with the selection region defined by Eq.~\ref{eq:color-color}.  There are 284 objects that meet the criteria defined by Eqns.\ ~\ref{eq:color-color} and \ref{eq:magcut}.  All 284 objects are detected at high confidence in the \Bhdf, \Vhdf, and \Ihdf\ bands. All but nine are detected in \Jhdf\ and all but two in \Hhdf, with all 284 being detected in either \Jhdf\ or \Hhdf.

Next, the pure color-color selection is followed by a photometric redshift cut based on photometric redshifts calculated from \Uhdf-through-\Hhdf\ spectral energy distributions as described in \S~\ref{sec:SEDfitting}. This photometric redshift cut requires that
\begin{equation}\label{eq:zphotcut}
1.8 \leq z_{phot} \leq 2.6, 
\end{equation}
where $z_{phot}$ is the redshift of the most likely model (i.e., the model with the lowest \chisq).  The lower photo-$z$ cut in Eq.~\ref{eq:zphotcut} is applied to eliminate potential low-$z$ interlopers, and is similar to the spectroscopic rejection of low-$z$ galaxies from samples such as those used by Shapley et al.\ (2005) or Steidel et al.\ (2004).  This 1.8$\leq$\zphot\ cut eliminates 87 objects out of the original 284, or $\sim$30\% of that total. As the photometric redshift distribution suggests (inset diagram in the right-hand panel of Fig.~\ref{sampleselection.fig}), some thirty of these objects are likely to be at $z$$>$1, but they nevertheless are eliminated from further analysis to avoid complications that could be introduced by a wide sample $\Delta$$z$ that spans a very wide range of cosmic time. The upper-$z$ cut, \zphot$\leq$2.6, is applied to avoid objects that are at too high a redshift to have enough rest-frame spectral coverage for accurate SED fitting. This upper cut is particularly important since it is impossible to estimate stellar population ages and galaxy stellar masses without a filter redward of the 4000\AA\ break, as would be the case at \z$\ga$2.6 for the \hdffilters\ filter set.  This \zphot$\leq$2.6 cut eliminates 106 higher-z objects, giving a final sample of 91 objects whose mean redshift is $<$\zphot$>$=2.3.  

It should be noted that the $z_{phot} \leq 2.6$ photometric redshift cut could potentially introduce a systematic bias since, in the absence of a visible Balmer/4000\AA\ break, 
extremely young $z_{phot} \leq 2.6$ galaxies could be confused for $z_{phot} > 2.6$ objects 
and thus eliminated from the sample; however, for a $z_{phot} \leq 2.6$ galaxy to lack a detectable break it would have to be extremely young and so only a small number of objects are likely to be missed in this way.  The reverse process --- that of higher-$z$ objects masquerading a extremely young $z$$\sim$2 stellar populations is unlikely to be important in practice given the very small number of $z$$\sim$2 objects with very young stellar populations that is observed in the sample in practice (\S~\ref{sec:ages}). 

In summary, the final sample of 91 objects resembles the selection and other properties (other than the luminosities) of the brighter Steidel et al.\ BX samples and thus henceforth will be referred to as the "HDF-BX sample".  While the focus of this study is on the \zphot-restricted HDF-BX sample of 91 objects, the analysis of the full, unrestricted sample of 284 objects show that the conclusions reached with the HDF-BX  sample are not significantly altered when using the full, unrestricted sample.

\subsection{Absolute rest-frame UV magnitudes}

\begin{figure}
\includegraphics[width=8cm]{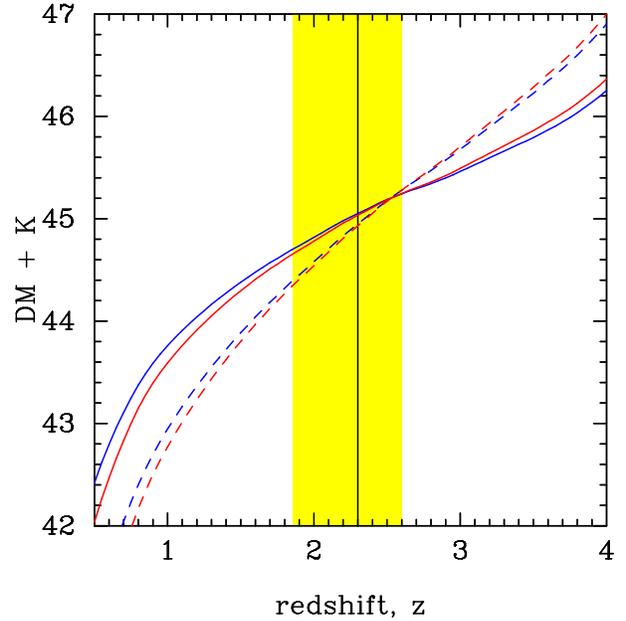}
\caption{
The cosmological corrections to calculating the absolute rest-frame 1700\AA\ magnitudes from apparent \Vhdf\ magnitudes.  Several different assumed SEDs are shown:  blue curves identify 10~Myr-old constant-SFR models while red curves indicate 100~Myr constant-SFR models; dashed curves show SEDs with no dust, while continuous curves show SEDs with \ebv=0.3. The redshift range of the present sample is shown as a shaded region, with the vertical line indicating the median photometric redshift.  In the present work a uniform correction is adopted for objects with $DM+K$=44.9. 
}
\label{fig.Kcorr} 
\end{figure}

The absolute rest-frame UV magnitude, $M_{1700}$, is calculated at the rest-frame wavelength of $\sim$1700, which matches the rest-frame wavelength of a  number of  LBG luminosity function studies (e.g., Steidel et al.\ 1999; Sawicki \& Thompson 2006a; Reddy \& Steidel 2009) and also approximately corresponds to the central wavelength of the \Bhdf\ bandpass at $z$=2.3.  It is derived using the usual cosmological distance modulus, $DM$, and $k$-correction, $K$,
\begin{equation}\label{eq:DM}
M_{1700} = m_{\lambda_{obs}} - DM - K,
\end{equation}
which can be rewritten as 
\begin{eqnarray}\label{eq:m2M}
M_{1700} & = & m_{\lambda_{obs}} - 5 \log (D_L / 10 {\rm pc}) + 2.5 \log (1+z) \nonumber \\
         &   & + (m_{1700}- m_{\lambda_{obs}/(1+z)}).
\end{eqnarray}
Here, $D_L$ is the luminosity distance and $m_{\lambda_{obs}}$ is the observed \Vhdf\ magnitude.  The last term of Eq.~\ref{eq:m2M}, $(m_{1700}-m_{\lambda_{obs}/(1+z)})$, is the $k$-correction color between rest-frame 1700\AA\ and the \Vhdf\ filter  in the {\it rest-frame} (e.g., Lilly et al.\ 1995; Sawicki \& Thompson 2006a). 

The $k$-correction color is expected to be very small because at \zs2.3 the \Vhdf\ filter probes close to rest-frame 1700\AA, as  is illustrated in Fig.~\ref{fig.Kcorr}. Because the change in $DM+K$ (Eq.~\ref{eq:m2M}) is small over the redshift range of the selection window (Fig.~\ref{fig.Kcorr}), rather than computing a separate correction for each galaxy based on its photometric redshift, a common value of $DM+K$=44.9 is assumed for all objects in the sample.  This approach introduces only small errors ($\sim \pm 0.4$ mag) in the derived absolute magnitudes, errors that are of the same order as those that would have been introduced by propagating the photometric-redshift uncertainties ($\Delta z_{phot}\sim$ 0.3 at these redshifts:  Sawicki et al.\ 1997,  Hogg et al.\ 1998) through Eq.~\ref{eq:m2M}.

\subsection{Comparison with other samples  at \zs 2}

\begin{figure*}
\begin{center}
\includegraphics[width=16cm]{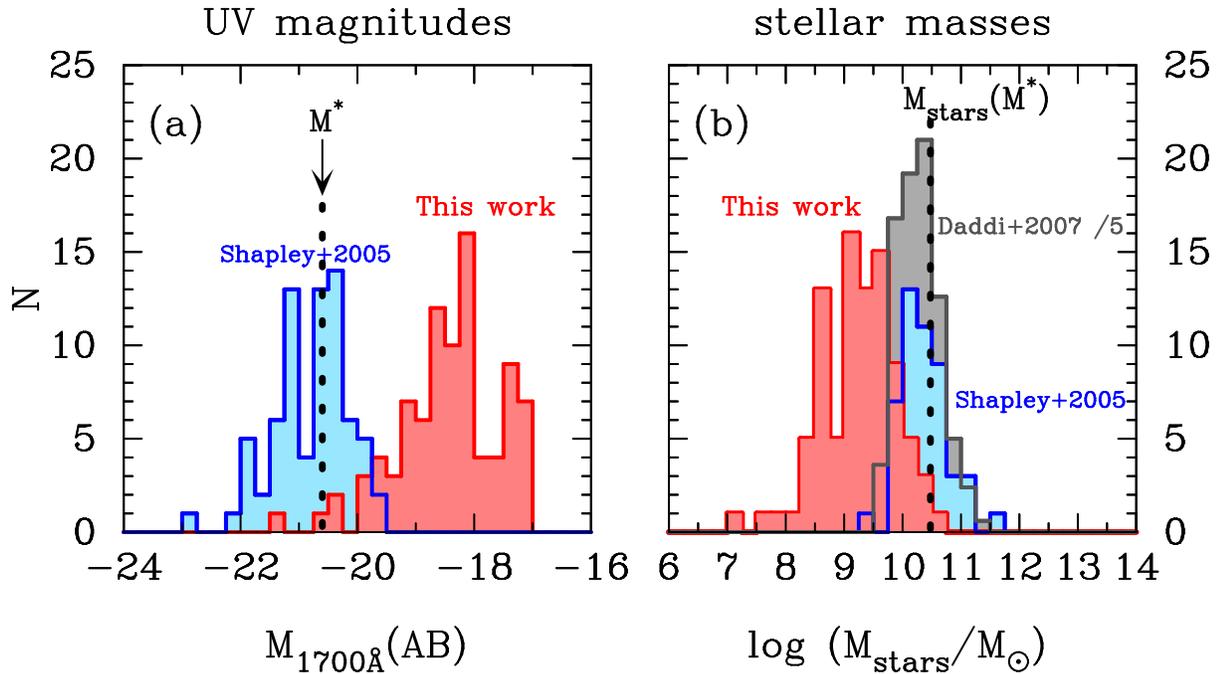}
\end{center}
\caption{
Comparison of the present HDF-BX sample with the BX sample
of Shapley et al.\ (2005) and the sample of Daddi et al.\ (2007).  Left panel: the Shapley et al.\ sample focuses on galaxies around the characteristic magnitude \Mstar, while the HDF-BX sample probes significantly fainter galaxies.  Right panel: stellar masses of galaxies in the present sample are typically an order of magnitude smaller than those in the Shapley et al.\ and Daddi et al.\ samples. Note that the Daddi et al.\ numbers have been scaled down by a factor of five to allow them to be displayed on this plot.}
\label{M-R-histograms.fig} 
\end{figure*}

The HDF-BX sample selection procedure described in \S~\ref{sec:the_sample} mimics the procedure used for selecting the \zs2 BX sample of Shapley et al.\ (2005). In both samples objects are rest-frame UV-selected and then \zs2 subsamples are identified on the basis of very similar color-color cuts.  The photometric redshift cuts imposed here on the HDF-BX sample (1.8$\leq$\zphot$\leq$2.6) mimic the spectroscopic cuts (1.7$<$$z_{spec}$$<$2.8) applied by Shapley et al.\ to their sample: in both samples the upper redshift cut eliminates $z$$\sim$3 LBGs, and the lower one rejects foreground interlopers. The resultant mean redshifts are also similar: $<$\zphot$>$=2.3$\pm$0.2 for the HDF-BX sample closely matches the $<$$z_{spec}$$>$=2.26$\pm$0.3 of the BX sample of Shapley et al.\ (2005). As is illustrated in the left panel of Fig.~\ref{M-R-histograms.fig}, the key difference between the samples lies in the fact that the HDF-BX sample consists of galaxies that are significantly fainter than those in the BX samples of Shapley et al.\ (2005), Steidel et al.\ (2004), and Erb et al.\ (2006): while their samples have a limiting magnitude of \R=25.5, the HDF-BX sample reaches to \Rave=28.0,  and while typical galaxies in the Shapley et al.\ (2005) BX sample have luminosities around L*, those in the HDF-BX sample are a factor of $\sim$10 fainter. 
 
The right panel of Fig.~\ref{M-R-histograms.fig} compares the stellar masses of the present HDF-BX sample (obtained through SED fitting described in \S~\ref{sec:SEDfitting})  with the similarly-selected but brighter BX sample of Shapley et al.\ (2005) and with the BzK galaxies of Daddi et al.\ (2007).  The stellar masses of the HDF galaxies are significantly lower than those in the other two samples: the median stellar mass in the HDF-BX sample is, at \s2.5$\times$10$^9$\Msun, \s10 times smaller than that in the BX sample of Shapley et al.\ (2005) and in the BzK sample of Daddi et al.\ (2007).  The HDF sample is  thus probing a similar but significantly fainter and apparently less massive UV-selected, star-forming population.

\section{SPECTRAL ENERGY DISTRIBUTION FITTING}\label{sec:SEDfitting}

\subsection{The technique}

Broadband spectral energy distribution fitting has in recent years become a popular tool for studying the properties of high redshift galaxies.   It was first used for that purpose by Sawicki \& Yee (1998) who used it to study the properties of the first samples of spectroscopically-confirmed high-$z$ galaxies in the HDF and was subsequently employed in numerous other studies of distant galaxies (e.g., Hall et al.\ 2001;  Sawicki 2001; Shapley et al.\ 2001, 2005; Papovich et al.\ 2001;  F\"orster Schreiber et al.\ (2004); Iwata, Inoue, \& Burgarella 2005; Yabe et al.\ 2009; Yuma et al.\ 2010, 2011; and many others); it now forms a part of the standard toolkit for high-$z$ galaxy research.  In the present work, the technique is applied to the HDF sample of very faint, sub-$L^*$ LBGs in order to explore the dependence of galaxy properties on UV luminosity. 

The SED-fitting approach quantitatively compares the photometry of a galaxy to a grid of models that spans a range of parameters of interest.  The comparison is done by means of a maximum likelihood test (\chisq\ fitting) and returns the best-fitting model parameters along with their uncertainties.  The technique is now past its infancy so only a brief description is provided here; for more details the reader is referred to papers by, e.g., Sawicki \& Yee (1998), Papovich et al.\ (2001), Shapley et al.\ (2001), and to the paper on SEDfit (Sawicki 2011), which is the publically-available SED-fitting software tool used in the work presented here. 

The generation of the model broadband SED grid begins with model galaxy spectra. In the present work, the underlying spectra are drawn from spectral synthesis models of  Bruzual \& Charlot (2003) with solar metallicity, the Salpeter (1955) stellar intial mass function (IMF) and a constant star formation rate. These spectra are first reddened using the starburst extinction curve of Calzetti et al.\ (2000), with the amount of reddening parametrized by the color excess \ebv.  Next, to simulate galaxies at high redshift, the reddened spectra are shifted in both wavelenght and intensity as dictated by the adopted cosmology of \OmOlHo\ = \cosmoparams.  Then, to include the effects of the absorption of UV photons by intergalactic neutral hydrogen along the line of sight, the spectra are attenuated below the Lyman limit using the prescription of Madau (1995).  Finally, the spectra are integrated through WFPC2 and NICMOS filter profiles to produce a grid of model fluxes to be compared against the data.  This grid is in effect 4-dimensional as it contains the following parameters: redshift, time since the onset of the dominant star formation (``age"), color excess \ebv, along with normalization of the model fluxes with respect to the data.  Table~\ref{tab:fit-parameters} lists the model grid parameters and their ranges.   Note that the input stellar evolutionary models, dust law, etc., used here to fit the HDF-BX objects are identical to those used by Shapley et al.\ (2005) for their BX sample. This commonality allows the two samples to be combined in order to cover a large luminosity baseline.

For each of the 284 objects in the color-selected catalog the photometry data are compared against the model grid and the best-fitting model is determined by a maximum likelihood (minimum \chisq) search.  In addition to the best-fitting parameters, the SEDfit software also provides estimates of their associated uncertainties by propagating the photometric uncertainties through the fitting procedure and deriving associated error volumes in the model grids.  The parameters derived from the fitting are:  age (i.e., time since the onset of the present episode of star formation), accumulated stellar mass, amount of reddening by dust, intrinsic (dust-corrected) star formation rate, and redshift.  Not all of these parameters are independent (e.g., accumulated stellar mass and star formation rate are both linked to the overall flux normalization of the model to the data), but all have their best-fitting values reported by SEDfit along with their associated uncertainties. 

SEDfit is capable of generating and fitting model SEDs based on a far wider range of star formation histories, stellar IMFs, dust attenuation laws, etc., than used in the present project, but the restricted  set of free parameters was a deliberate choice made to allow the comparison of the present sample with the luminous BX sample of Shapley et al.\ (2005).  The dependence of fit results on the assumed metallicity, star formation history, and the like have been explored previously by Sawicki \& Yee (1998), Papovich et al.\ (2001), and many others, and are reasonably well understood. For example, it is known that SED-fitting cannot usefully constrain metallicity or star formation history --- but, by the same token, metallicity differences in the models do not impact the resulting fit parameters for other quantities such as stellar mass and SFR.  Galaxy stellar masses derived via SED fitting are reasonably robust to the various model assumptions with typical systematic uncertainties of factors of 2--3. Star formation rates, on the other hand, are highly sensitive to the unknown star formation histories. This paper proceeds on the assumption that star formation rates are constant for galaxies of all UV luminosities.  This seems a reasonable assumption in light of the tightness of the SFR--$M_{stars}$ relation (\S~\ref{sec:mass-sfr}); potential systematic uncertainties are further mitigated by the approach of relative --- rather than absolute --- comparisons between galaxies of different luminosities.

\begin{table*} 
\label{tab:fit-parameters}
\begin{center}
\caption[]{Model Parameters.  The first four are free parameters of the fit, the last four are fixed.}
\begin{tabular}{lccl}
\hline
parameter &
value(s) & 
step size & 
remarks\\
\hline
redshift & 0 -- 5 & 0.05\\
log (age/Gyr) & 5.1 -- 10.3 & 0.1 &  \\
$E(B-V)$ & 0 -- 0.6 & 0.02 & Calzetti et al.\ (2000) \\
flux normalization & any & exact & see eq.\ 3 of Sawicki (2002) \\
\hline
star formation history & constant SFR & --- & Bruzual \& Charlot (2003)\\
IMF & Salpeter (1955) & ---  & $ 0.1 < M/M_\odot< 100$ \\
metallicity & 1 $Z_\odot$ & --- & \\
intergalactic attenuation & 1 & --- & Madau (1995) \\
\hline
\end{tabular}
\end{center}
\end{table*}

\subsection{SED fitting results and examples}

\begin{figure*}
\begin{center}
\includegraphics[width=16cm]{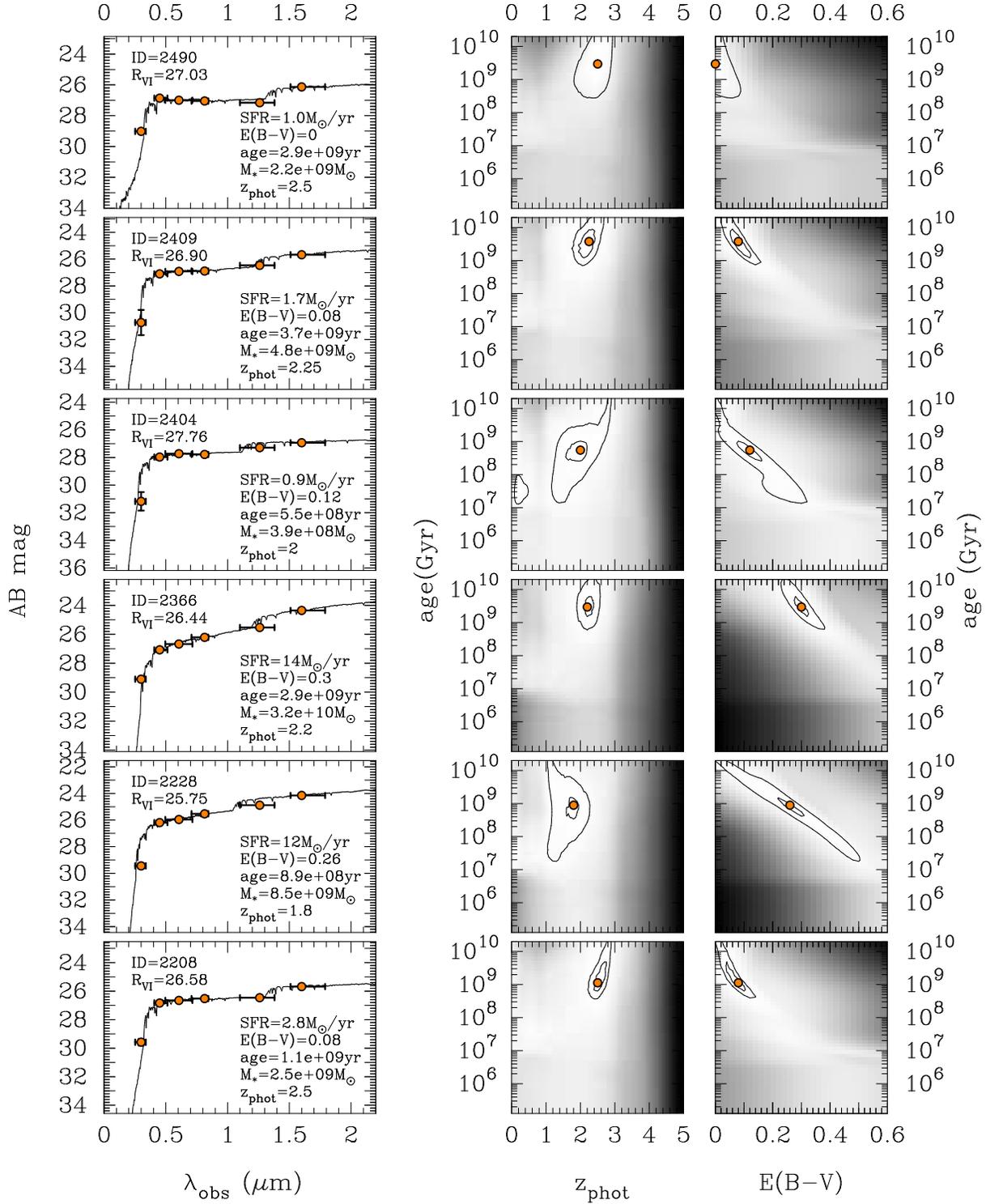}
\end{center}
\caption[]{
Examples of SED fits.  In the left-hand panels, the horizontal error bars are not uncertainties but rather denote the FWHM range of the filter transmission curves.  Note that the \Uhdf-band (shortest $\lambda$) photometric points were not used in the SED fitting and are shown here for completeness only.  In the middle and right panels, the inner contours are drawn at surfaces of constant \chisq\ that correspond to 68\% statistical significance, while the outer contours correspond to 95\% confidence.   
}
\label{fig.fitexamples} 
\end{figure*}

Representative examples of the fits are shown in Fig.~\ref{fig.fitexamples}. The left-hand panels of Fig.~\ref{fig.fitexamples} compare  the photometric data (points) with the Bruzual \& Charlot (2003) spectra that give the best-fitting model SEDs.  The FWHM of the filter transmission curves are shown as horizontal errorbars, but it should be kept in mind that these are not actual errorbars in the normal understanding of that term, but rather are meant to give an indication of the wavelength range over which the observed and model spectra are integrated when producing broadband magnitudes.

The contour plot panels of Fig.~\ref{fig.fitexamples} show the values of the best-fit redshift, age, and color excess \ebv, together with the permissible ranges of these parameters.  The inner contours (in some cases largely hidden under the best-fit point) correspond to a probability of 68.3\% that the observed data were drawn from the enclosed models; the outer contours show the 95\% probability.  The error contours were calculated for each object by perturbing the observed photometry assuming Gaussian uncertainties and then refitting the perturbed data.  Two hundred such Monte Carlo re-fits were performed for each object and on the basis of these 91$\times$200=18,200 realizations, the values of \chisq\ that enclose the desired percentage probabilities were estimated for plotting. 

The redshift error contours in Fig.~\ref{fig.fitexamples} are remarkably tight and not particularly degenerate with the other parameters.  This fact gives confidence that the lack of spectroscopic redshifts for the very faint objects studied here is not a serious impediment. The error contours on the other parameters are also well behaved and although some degeneracies exist between age and extinction, characterizing the sizes of the errorbars through this Monte Carlo-based procedure makes it possible to proceed with the fit results to help us understand the nature of this sample of very faint galaxies.

\section{RESULTS}\label{sec:results}

\subsection{Preliminaries}\label{sec:results:preliminaries}

This section discusses some interesting trends found in the properties of UV-selected galaxies at \zs2.  In several instances the sample of 91 very faint HDF objects is compared against similarly-selected and similarly-fit but more luminous galaxies from the BX  sample of Shapley et al.\ (2005).  In other instances, that Shapley et al.\ (2005) sample is used to augment the HDF sample in order to extend or highlight important trends present in the data as a function of galaxy luminosity.   

The exact values of the best-fitting parameters can depend sensitively on the details of the assumed dust extinction law, star formation history, stellar initial mass function, etc. (see, e.g., Sawicki \& Yee 1998, Papovich et al.\ 2001) so it is vital to use models with exactly the same parameters when comparing different samples. With that thought in mind, the SED fits to the present HDF-BX sample use exactly the same underlying models as those for the brighter BX sample of Shapley et al.\ (2005): solar metallicity, constant SFR Bruzual \& Charlot (2003) models with the Salpeter (1955) IMF and Calzetti et al.\ (2000) dust. Consequently,  the two samples can be regarded as directly comparable. 

Notwithstanding the commonality of models, one of the strongest limitations of broadband SED fitting is the need to assume a star formation history, which cannot be reliably constrained for a galaxy from the small number of photometric points available. Whereas many parameters of interest --- such as reddening or stellar masses --- are relatively insensitive to the assumed star formation history (e.g., Sawicki \& Yee 1998; Papovich et al.\ 2001), stellar population ages and star formation rates are highly dependent on the assumed star formation history.  Additionally, prior star-forming episodes may well have occurred but their existence in a galaxy can be masked by the glare of the newly-born massive stars belonging to the most recent star-forming episode; furthermore, star formation in high-redshift galaxies may in fact be increasing with time (e.g., Lee et al.\ 2010; see also \S~\ref{sec:limitations}) thereby effectively masking the presence of older stars. For these two reasons --- the challenge of constraining the history of even the most recent episode of star formation along with difficulty of detecting possible earlier star-forming episodes --- interpreting the ages derived from SED fitting can be problematic. 

Neglecting possible prior star-forming episodes and considering only a single episode of star formation, constant star formation models give older ages than do models in which star formation declines with time.  For this reason, constant SFR fits that yield ages older than that of the Universe can be regarded as likely having declining star formation histories as opposed to the constant SFR ones that were assumed in the fitting.  A number of such objects are present in both the HDF-BX and the Shapley et al.\ (2005) samples and these objects --- while plotted (in red) in the Figures that follow --- are excluded from the main of the analysis.  On the other extreme, objects with extremely young best-fit ages are also problematic: such an object is unlikely to have formed immediately prior to the epoch of observation and is, instead, probably a normal galaxy whose light is dominated by a very  recent burst of star formation that masks an older, underlying stellar population.  Only a very small handful of such objects exist in the HDF-BX sample, with a somewhat larger fraction present in the Shapley et al.\ (2005) sample.  Since the goal of the present work is to establish longer-term behavior of galaxies, such "young" galaxies (defined here as best-fit age $<$ 30  Myr) are thus also treated with suspicion and --- while shown (in blue) in the Figures that follow --- are excluded from the main of the analysis. 

Because of these issues, stellar population ages determined from the SED fitting are not analyzed in detail in what follows. With these preliminaries out of the way, let us turn to some of the interesting trends found in the data.

\subsection{Dust}\label{sec:dust}

It has now been known for some time now that UV-selected high-$z$ galaxies posses significant amounts of interstellar dust, (e.g., Meurer et al.\ 1997; Sawicki \& Yee 1998; Papovich et al.\ 2001; Vijh, Witt, \& Gordon 2003; Reddy et al.\ 2006).  However, the effect of dust remains unknown in very faint and/or low-mass galaxies at high redshift. This section examines the dependence of dust attenuation on UV luminosity, reaching very deep into the galaxy population, by pushing to \Rave=28, or 3.5 magnitudes below \Mstar at \zs2. 

\begin{figure*}
\begin{center}
\includegraphics[width=16cm]{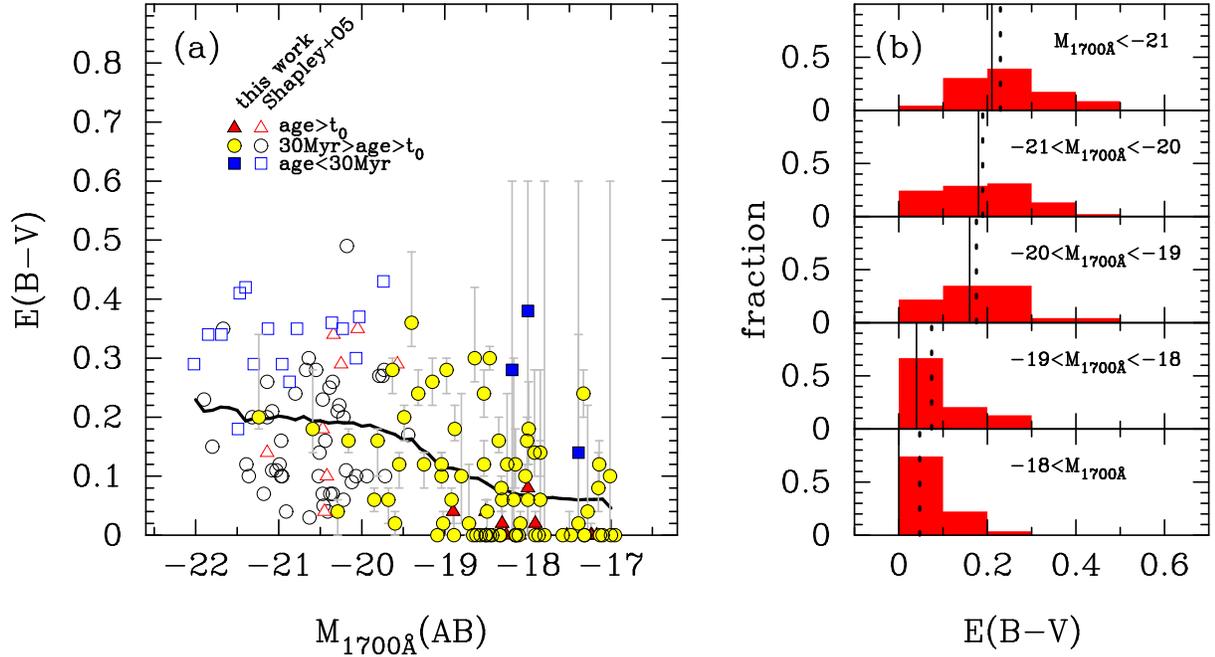}
\end{center}
\caption{
Dependence of extinction on UV luminosity. The left panel shows color excess values as a function of absolute UV magnitude (not corrected for extinction) with objects in different best-fit ages indicated by different symbols. The black line shows the running mean \ebv\ value, taken with a 2 mag-wide box. A systematic reduction in extinction towards fainter rest-frame UV luminosity can be seen.  Note that no uncertainties are shows for the Shapley et al. (2005) data because these authors do not report them; however, their uncertainties are likely significantly larger than those at similar luminosities in the HDF. The right panel shows the distributions of color excess values for objects grouped by different UV magnitudes, with both the present HDF and the Shapley et al.\ (2005) samples treated jointly. The solid and dashed vertical lines in each panel mark the median and mean color excess values, respectively.  Again, a decrease in the amount of extinction can be seen with decreasing UV luminosity. 
}
\label{ebv-vs-mag.fig} 
\end{figure*}

\begin{table*} 
\label{tab:dust_results}
\begin{center}
\caption[]{Extinction as a function of UV magnitude.  Uncertainties given are simple RMS measurements.  Stellar masses are estimated using the \Muv -\Mstars\ relation given by eq.~\ref{eq:mass-mag}.}
\begin{tabular}{lcccc}
\hline
UV magnitude  &
log ($M_{stars}$/\Msun) & 
median E(B--V)  &
mean E(B--V) &
mean $A_{1700}$ \\
\hline
\Muv $<$ $-21$                  & $>$10.60 & 0.21 & 0.23 $\pm$ 0.10 & 2.24 $\pm$ 0.98\\
$-21$ $<$ \Muv $<$ $-20$ & 10.35       & 0.18 & 0.19 $\pm$ 0.11 & 1.84 $\pm$ 1.07\\
$-20$ $<$ \Muv $<$ $-19$ &   9.84       & 0.16 & 0.18 $\pm$ 0.11 & 1.75 $\pm$ 1.07\\
$-19$ $<$ \Muv $<$ $-18$ &   9.33       & 0.04 & 0.07 $\pm$ 0.09 & 0.68 $\pm$ 0.88\\
$-18$ $<$ \Muv                  &   $<$9.07       & 0    & 0.04 $\pm$ 0.07 & 0.39 $\pm$ 0.68\\
\hline
\end{tabular}
\end{center}
\end{table*}

Figure~\ref{ebv-vs-mag.fig}(a) shows the dependence of color excess, \ebv, on galaxy UV luminosity. The data suggest that fainter BX galaxies suffer less extinction than the more luminous ones, a trend that extends to the very faint limit of the HDF data. This finding is in agreement with the conclusion of Adelberger \& Steidel (2000), but here, for the first time, the trend is extended to high-$z$ galaxies $\sim$3.5 magnitudes fainter than \Mstar.  The reality of this trend is verified by Monte Carlo simulations which suggest that moderately dusty galaxies (\ebv$\sim$0.15) are unlikely to be scattered by photometric errors to look like the nearly dust-free objects seen at the faint end of Fig.~\ref{ebv-vs-mag.fig}. Figure~\ref{ebv-vs-mag.fig}(a) may also appear to show a preference for galaxies dominated by very young stellar populations (blue squares) to be more dusty than the mean, especially so at the bright end of the population where the Shapley et al.\ (2005) data dominate.  However, at least within the HDF-BX sample, Monte Carlo simulations suggest that this effect is an artifact of correlated extinction-age errors, and this may also be the case for the Shapley et al.\ data given their large photometric uncertainties.  In summary, the tendency for low-luminosity BX galaxies to have little dust is likely real but any apparent trends of dust with age are not. 

Figure~\ref{ebv-vs-mag.fig}(b) quantifies the dependence of extinction on UV brightness by showing the distribution of \ebv\ values in bins of galaxy UV magnitude. As in the left-hand panel of Fig.~\ref{ebv-vs-mag.fig}, there is a trend for fainter galaxies to have lower \ebv\ values, and both the median (solid vertical lines) and mean (dashed lines) \ebv\ values decrease for the fainter magnitude bins.  These results are summarized in Table~\ref{tab:dust_results}.  Inspecting the $A_{1700}$ values in Table~\ref{tab:dust_results}, it is interesting to note that while only about one out of five UV photons escapes from an \Lstar\ BX galaxy, sub-\Lstar\ galaxies are far more ``naked'':  in the mean we are able to see more than half of the UV photons produced by \Mstar+3 BX objects.  One useful consequence of this low dust attenuation is that sub-\Lstar\ galaxies may be simpler to study than their \Lstar\ counterparts. 

As is discussed in \S~\ref{mass-vs-lum:sec}, there is a correlation between stellar masses and luminosities of BX galaxies, with more luminous objects being more massive.  Given the luminosity-extinction relation discussed above, it is thus clear that the low-mass BX galaxies contain smaller amounts of dust than their higher-mass $\sim$\Lstar\ cousins.  It is plausible that the low-mass galaxies have lower baryon densities than the more massive objects, or, alternatively, that they are physically smaller. Either of these two possibilities is consistent with the luminosity/mass vs.\ reddening trend. Whatever the reason, the UV photons produced by their stars have smaller dust column densities to risk on their way into the intergalactic medium.  Some of the issues related to the dust-stellar mass correlation will be explored further in \S~\ref{sec:dust-mass}.

\subsection{Stellar mass-UV luminosity relation}\label{mass-vs-lum:sec}

\begin{figure}
\includegraphics[width=8cm]{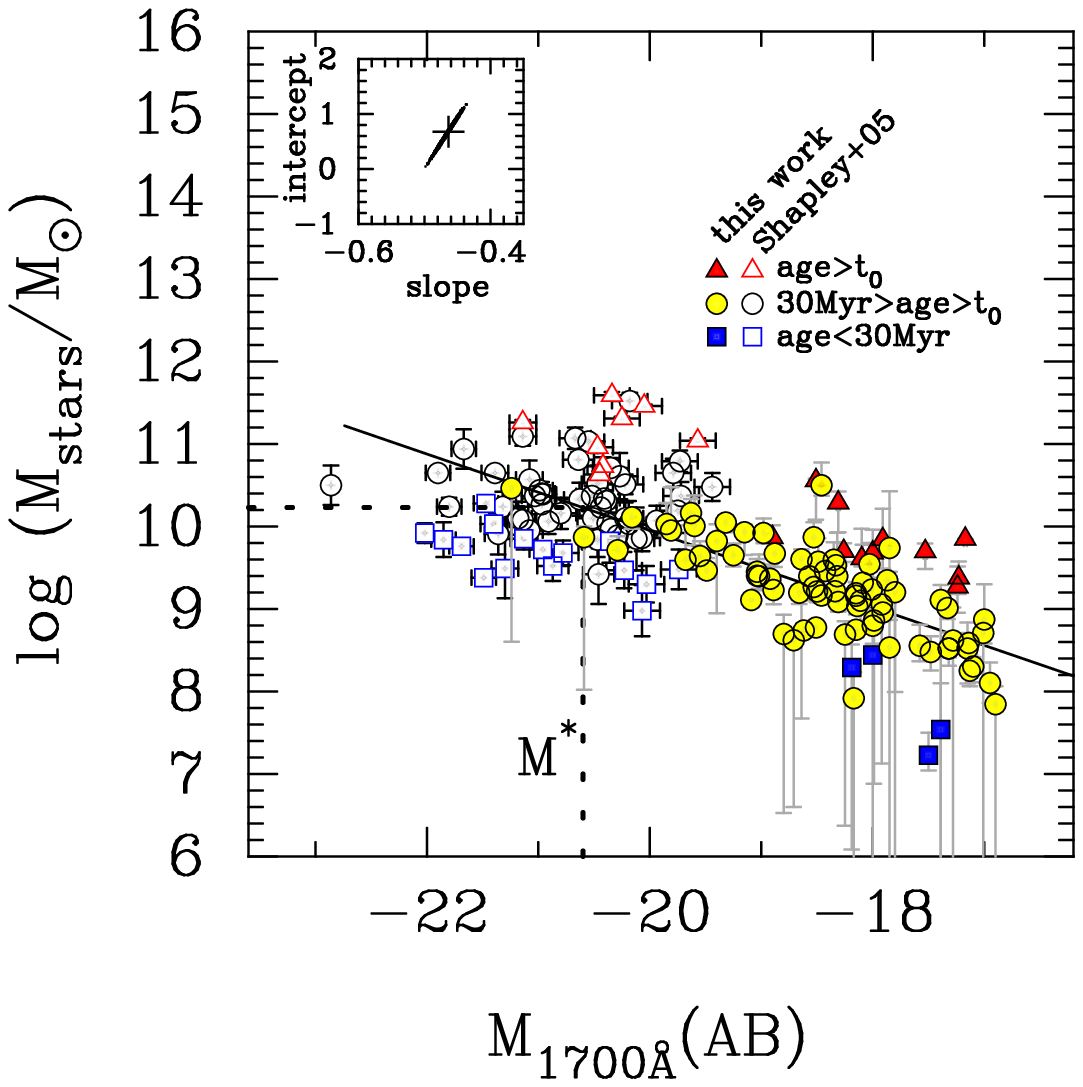}
\caption{
Stellar mass as a function of UV magnitude.  The stellar masses were calculated using Eq.~\ref{eq:mass-mag}  and UV magnitudes are {\emph not} corrected for dust}. Different symbol shapes show objects in different age ranges (constant star formation rate histories have been assumed for all objects).  Objects marked with triangles are older than the approximate age of the Universe at \z=2.3.   The 1-$\sigma$ error contour for the best-fit line is shown in the inset.

\label{Mstars-vs-mag.fig} 
\end{figure}

Given that the UV light from faint BX galaxies appears to be fairly unobscured (\S~\ref{sec:dust}), it is interesting to ask how stellar mass correlates with the emergent UV luminosity in these objects. With this in mind, Fig.~\ref{Mstars-vs-mag.fig} examines the dependence of stellar mass on UV magnitude (not corrected for dust) and shows that there is a strong correlation between stellar mass and UV luminosity. This correlation is unlikely to be an artifact of an observational bias:  concentrating on the HDF-BX sample, where the data cover a relative large luminosity baseline, one does not see any luminous (\Muv$\sim-20$) under-massive ($10^{8-9}$\Msun) galaxies, although the data are deep enough to identify such objects were they present in significant numbers; similarly, not many over-massive galaxies are found at the faint end of the population, though such galaxies, too, would be easily detected if present in significant numbers.  The \Muv - \Mstars\ relation in Fig.~\ref{Mstars-vs-mag.fig} thus appears to be real. 

The relation in Fig.~\ref{Mstars-vs-mag.fig} can be fit as 
\begin{equation}\label{eq:mass-mag}
\log \left( \frac {M_{stars}}{M_\odot} \right) = (0.68 \pm 0.49) - (0.46 \pm 0.03) M_{1700}
\end{equation}
(black line in Fig.~\ref{Mstars-vs-mag.fig})\footnote{ Note that the values in Eq.~\ref{eq:mass-mag} here are slightly different from those reported earlier in Sawicki et al.\ (2007) and subsequently used by others (e.g., Reddy \& Steidel 2009) --- these differences are due to the updated, deeper sample that's used in the present work. }.  One important application of this relation is that it can be used to estimate BX galaxy stellar masses from optical (rest-frame UV) data alone without the need to resort to infrared observations. The uncertainties in the slope and intercept values given in Eq.~\ref{eq:mass-mag} are strongly correlated (see inset panel in Fig.~\ref{Mstars-vs-mag.fig}), which means that the conversion between rest-frame UV magnitude and stellar mass is actually quite tight --- much tighter than the one-dimensional uncertainties might imply at first glance.

Figure~\ref{Mstars-vs-mag.fig} suggests that the  \Muv-\Mstars\ relation is valid from \Mstar\ to \Mstar+4, or over \Muv$=-21$ --$-17$.  It is not clear whether at brighter magnitudes, \Muv$<$\Mstar, the relation holds given the apparent flattening of the data at \Muv $<$ \Mstar, the lack of data beyond \Muv=$-$22, and  the expectation that different physics play a role above the knee of the LF. Although no data exist at \Muv$>-17$, it is plausible that the relation extrapolates well to fainter galaxies.

The results of Fig.~\ref{Mstars-vs-mag.fig} and Eq.~\ref{eq:mass-mag} are robust in that they do not change dramatically if HDF BX objects outside the $1.8\leq z_{phot} \leq 2.6$ range are included: the scatter in the relation increases and the slope of the relation steepens somewhat, but these changes are not unexpected since the stellar masses are then poorly constrained, particularly at \zphot$>$2.6. It is therefore preferable to use only the restricted $1.8\leq z_{phot} \leq 2.6$ HDF-BX sample, as is done through the rest of this paper. 

It should be kept in mind that the galaxies here are UV-selected, so the data in Fig.~\ref{Mstars-vs-mag.fig} are certainly not representative of the entire \zs2 galaxy population as they could miss, for example, quiescent galaxies, or very heavily dust-obscured ones.  However, when applied to UV-selected samples, the observed mass-luminosity correlation gives an interesting tool for studying galaxy evolution: for example, it can link UV-selected samples from different studies (e.g., Savoy, et al.\ 2011), or provide a way to convert deep LFs into mass functions (\S~\ref{sec:updatedLF}; also Reddy \& Steidel 2009). It also hints at an underlying correlation between SFR and stellar mass (\S~\ref{sec:mass-sfr}). It is thus worth examining in more detail. 

The RMS scatter in the \Muv-\Mstars\ relation for the combined, Shapley + HDF-BX, sample is $\sim$0.4 in $\log$\Mstars.  Within the HDF-BX sample the scatter goes down from high values at the faint/low-mass end as one moves to the more luminous/massive galaxies. Monte Carlo simulations that perturb the photometry of the HDF-BX galaxies indicate that the observed scatter is entirely consistent with random photometric scatter from an intrinsically perfectly tight relation described by Eq.~\ref{eq:mass-mag}. 

The \Muv-\Mstars\ trend becomes clear only when the data are extended below \Lstar\ and may well be absent in the more luminous BX galaxies alone.  One explanation for the absence of correlation in the Shapley et al.\ (2005) sample is that large photometric scatter in their relatively shallow ground-based data is propagated into their stellar mass estimates.  An alternative explanation stems from the fact that Shapley et al's galaxies, which have higher masses and larger star formation rates, have a much larger intrinsic range of dust extinctions than do the fainter HDF objects (\S~\ref{sec:dust}).  If this is the case, then at the brighter magnitudes studied by Shapley et al.\ (2005) the stellar mass-UV luminosity correlation may be naturally broad, making in-the-mean trends difficult to discern.  In contrast, the fainter HDF-BX galaxies, which suffer less extinction, are less affected and so hold truer to the underlying correlation. In either case, the deep HDF data allow the luminosity-mass trend among sub-\Lstar\ BX galaxies to stand out.  

In Fig.~\ref{Mstars-vs-mag.fig}, the blue squares mark object whose best-fit ages are extremely young (younger than 30 Myr), while the red triangles flag objects whose best-fit ages are older than the age of the Universe. Note that at a given UV magnitude, these seemingly very young objects tend to have stellar masses that are low relative to the norm, while the objects with very old age fits have masses that are high.  This trend would at first glance appear consistent with the build-up of stellar mass with time in galaxies.  However, as tempting as this conclusion may appear, it may not be warranted given the random uncertainties due to photometric errors and the systematic ones due to assumptions about star formation histories that were discussed in \S~\ref{sec:results:preliminaries}.  

It is next interesting to consider what is the stellar mass of a "typical" (i.e., \Lstar) BX galaxy at \zs2.3.  To this end, the vertical dashed line in Fig.~\ref{Mstars-vs-mag.fig} indicates the location of the \Lstar\ "knee" of the BX galaxy luminosity function ($M^*_{1700}=-21.0^{+0.5}_{-0.6}$ at \zs2.3 --- \S~\ref{sec:updatedLF}; see also Sawicki \& Thompson 2006a, Reddy \& Steidel 2009); the corresponding characteristic stellar mass at the knee is \Mstars($L^*_{1700}$)=$(2.60^{+2.33}_{-1.08})\times 10^{10}$\Msun.  This characteristic stellar mass is a factor of several lower than the bimodality transition mass that separates low-$z$ galaxies into two distinct families ($\sim 6 \times 10^{10}$\Mstar; Kauffmann et al. 2003b after adjusting from their Kroupa (2001) IMF to the Salpeter IMF used here) or the characteristic mass in the local IR-selected galaxy luminosity function (Cole et al.\ 2001). 

\subsection{Ages and Star Formation Histories}\label{sec:ages}

What can we learn about the ages and star formation histories of sub-\Lstar\ galaxies from the results of SED fitting?  As was pointed out in \S~\ref{sec:results:preliminaries}, best-fit ages are highly dependent on the assumed star formation history and consequently it is impossible to say anything absolute about the ages of the star-forming episodes that we observe in the individual BX galaxies.  Moreover, correlated uncertainties in the best-fit parameters can make even relative age trends uncertain:  For example, Monte Carlo simulations that perturb the photometry of the HDF-BX galaxies confirm this view, showing that the apparent age trends seen in Fig.~\ref{Mstars-vs-mag.fig} are likely simply a result of photometric scatter. Naively, one might be tempted to conclude that the objects fit with very old ages have declining star formation histories and are thus over-massive for their star formation rates; and objects fit with very young ages are experiencing transient bursts of star formation.  However, Monte Carlo simulations show that the distribution seen in Fig.~\ref{Mstars-vs-mag.fig} can be produced by random photometric scatter of a uniform population in which all objects have identical, moderate ages and are forming stars at a fixed (though object-dependent) rate.  The apparent age trends in Fig.~\ref{Mstars-vs-mag.fig} are simply a result of correlated errors in best-fit parameters. 

On the positive side, the existence of the luminosity-stellar mass and SFR-stellar mass correlations (Fig. ~\ref{Mstars-vs-mag.fig}) suggests that star formation histories in \zs2 BX galaxies may be relatively smooth rather than rapidly fluctuating. Turning to the HDF-BX sample, where the photometric uncertainties are smaller and consequently so is random scatter in the best-fit parameters, the tightness of the correlation and the relative small number of outliers are impressive. Were the majority of the galaxies in the sample undergoing significant recent bursts of star formation, one would expect to see many more outliers than are observed.  This fact has three important consequences:  First, it validates the assumption of constant star formation histories that was made in the SED fitting (\S~\ref{sec:SEDfitting}).  Second, it suggests that by \zs2 many star-forming galaxies have been forming stars in a relatively steady mode for significant periods of time, rather than being dominated by frequent bursts or other fluctuations to their star formation history. Third, it suggests that there is not a large population of galaxies with very large stellar masses hidden by small amounts of new star formation.

In summary, while any age trends seen in the best fit results are unlikely to be real, the existence of the \Muv - \Mstars\ currelation suggests that star formation in sub-\Lstar\ UV-selected galaxies at \zs2 may be approximately constant rather than highly variable.

\begin{figure*}
\begin{center}
\includegraphics[width=12cm]{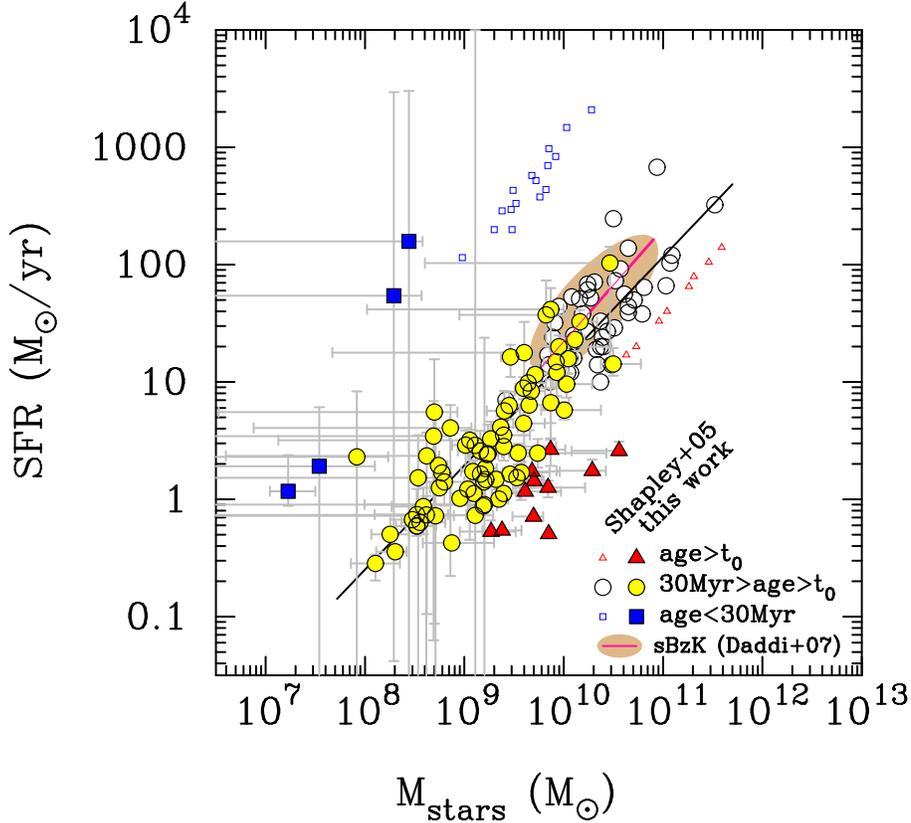}
\end{center}
\caption{
The dependence of SFR on stellar mass.  No errorbars are shown for the Shapley et al.\ (2005) data as these authors do not provide uncertainties on their SFR measurements.  The black line shows the fit to the combined HDF-BX and Shapley et al.\ BX populations, after the extreme-age outlier objects have been removed (see \S~\ref{sec:mass-sfr}).  The beige region and pink line show the locus of the the 24$\mu$m-detected sBzK galaxies of Daddi et al.\ (2007) and their fit to those data. 
}
\label{SFR-vs-Mstars.fig} 
\end{figure*}

\subsection{Relation between star formation rate and stellar mass}\label{sec:mass-sfr}

The correlation between rest-frame UV magnitude and mass (\S~\ref{mass-vs-lum:sec}; Fig,~\ref{Mstars-vs-mag.fig}) may seem unexpected at first, but it is not surprising given that --- for sub-\Lstar\ galaxies --- dust corrections are rather small:  unlike for the more massive objects, emergent UV luminosity is a reasonably good indicator of the star formation rate in a sub-\Lstar\ BX galaxy. Nevertheless, it is important and interesting to correct for dust attenuation, and such dust-corrected quantities are examined in this section. 

Figure~\ref{SFR-vs-Mstars.fig} shows the dependence of the intrinsic (dust-corrected) star formation rates on the galaxies' stellar masses. The present BX galaxy results improve on a similar relation reported in Sawicki et al.\ (2007) by extending it to lower still stellar masses. It is remarkable that a SFR-stellar mass correlation can be seen at \zs2.3 over as much as three orders of magnitude in stellar mass.

Also shown in Fig.~\ref{SFR-vs-Mstars.fig} is the location of the 24$\mu$m-detected star-forming BzK (i.e., sBzK) galaxies studied by Daddi et al.\ (2007; beige oval), along with their best-fit line to those data (pink line).  Both the Daddi et al.\ sBzK sample, and the Shapley et al.\ (2005) BX galaxies can be viewed as a high-mass extension of the HDF-BX galaxies studied in the present work.  

The Shapley et al.\ sample has two sizable outlier populations that are puzzling and likely unphysical artifacts:  the small red triangles are objects that were fit with ages older than the age of the Universe ($\sim$2.8 Gyr at \zs2.3), while the small blue squares mark objects with very young ages ($<$30 Myr). Each population lies on a distinct locus from the moderately-aged objects in their sample --- possibly an artifact of their fitting procedure.  In light of this, it might be preferable to use the Daddi et al.\ zBzK objects as a high-mass complement of the faint HDF-BX sample, but for the sake of commonality of selection techniques it was decided to use the Shapley et al.\ BX galaxies after rejecting their two outlier populations. 

Ignoring the two extreme-age outlier populations in both the Shapley et al.\ data and in the present HDF-BX sample, the BX galaxy SFR-stellar mass relationship is fit by
\begin{equation}\label{eq:sfr-mass} 
\log \left( \frac{SFR}{M_\odot yr^{-1}} \right) = (0.89 \pm 0.03) \log \left( \frac{M_{stars}}{M_\odot} \right) - (7.69 \pm 0.27),
\end{equation}
which is shown as the black line in Fig.~\ref{SFR-vs-Mstars.fig}.  
Note that using the sBzK galaxies instead of the BX galaxies of Shapley et al.\ would produce a similar fit to the overall population, with perhaps just a slightly steeper slope. 

Taken at face value, the correlation between stellar mass and SFR (Eq.~\ref{eq:sfr-mass}) suggests that the rate at which a galaxy converts gas into stars (i.e., its star formation rate) is dictated by the mass in stars that that galaxy has already assembled. This seems surprising at first since one might expect that, for a constant SFR, stellar mass should increase with time, and such an increase should make galaxies drift off of the relationship found above.  Although evidence for such a drift may be sought in the presence of the massive, old galaxies in Fig.~\ref{SFR-vs-Mstars.fig}, the fairly small number of such galaxies in the HDF-BX sample suggests that the constant-SFR scenario is unlikely.  Moreover, Monte Carlo simulations suggest that this massive old HDF-BX population can be produced by simple photometric scatter of a moderately-aged population of galaxies that lie on the relation given by Eq.~\ref{eq:sfr-mass}, thus further hinting that a different scenario may be at play. 
One such scenario, which posits that the galaxy's star formation rate is proportional to the dark matter accretion rate of its growing dark matter halo, is examined in \S~\ref{sec:sfr-mass-model}.

\section{IMPLICATIONS}\label{sec:discussion}

\subsection{Updated Luminosity Function and Cosmic SFR Density}\label{sec:updatedLF}

\begin{figure*}
\includegraphics[width=12cm]{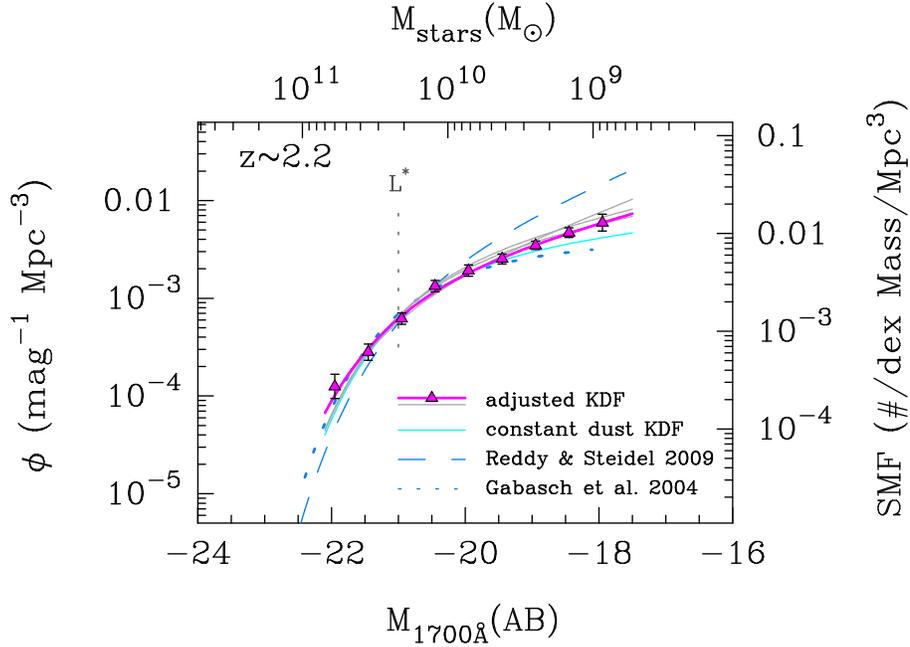}
\caption{
The UV-selected galaxy lumonosity function at \zs2.3.  Data points show the KDF (Sawicki \& Thompson 2006) result updated for the effects of magnitude-dependent attenuation. The heavy solid curve shows the Schechter fit to the data, while the lighter-colored curves are fits to other implementations of the \Veff\ correction (see Sawicki \& Thompson 2006a for details).  The Reddy \& Steidel (2009) and Gabasch et al. (2004) luminosity functions are shown for comparison.  Because of the tight correlation between UV luminosity and stellar mass (see \S~\ref{mass-vs-lum:sec}), the LF can be readily used to estimate the stellar mass function, shown here with the other top/right axes. 
}
\label{LF-MF.fig} 
\end{figure*}

One of the most basic yet important descriptors of a galaxy population is its luminosity function (LF) and much effort has gone into characterizing the shape and evolution of the LF of high-redshift star-forming galaxies (e.g., Sawicki, Lin, \& Yee 1997; Steidel et al.\ 1999; 
Sawicki \& Thompson 2006a; Iwata et al.\ 2007; Reddy \& Steidel 2009). This section revisits the faint end of the BX galaxy examined by Sawicki \& Thompson 2006a, in order to update their result in light of the effects that luminosity-dependent extinction can play on the estimation of effective survey volume and, hence, on the number density of galaxies.  It is important to be clear that this does not represent an attempt to correct the UV luminosities of galaxies for dust obscuration, but, rather, to better include the effect that dust has on the model galaxy colors needed to estimate the effective survey volume in the \Veff\ method of LF estimation. 

Estimation of the LF is complicated by issues such as incompleteness corrections and proper determination of the survey volume.  When dealing with color-selected samples such as BX or LBG galaxies, survey volume can depend on assumptions made about the intrinsic color distribution of high-redshift galaxies, which in turn can depend on the amounts of dust assumed to be present in these galaxies. In this respect, Sawicki \& Thompson (2006a) found that while LBG LFs at \zs 3 and 4 are not strongly sensitive to such underlying assumptions about galaxy SEDs, the \zs2.3 BX galaxy LF can be.  In particular, they pointed out that the assumed amount of reddening will affect the normalization of the BX galaxy LF (see their Figs.\ ~7 and 8) and examined the impact of different amounts of assumed reddening on their LF results. However, lacking concrete knowledge of extinction in sub-\Lstar\ BX galaxies, the scenarios they examined assumed luminosity-invariant \ebv=0.15. With the finding of luminosity-dependent extinction discussed in \S~\ref{sec:dust} it is now possible revisit this previous analysis and to better constrain the shape of the BX galaxy LF.   

The KDF LF analysis of Sawicki \& Thompson (2006a) used the effective volume (\Veff) method and their data are re-analysed here using the same approach but using magnitude-dependent \ebv\ values when constructing the mock galaxies used in gauging the redshift selection function and the effective volume of the survey, \Veff.  To this end the KDF fiducial dust-corrected 100-Myr starburst model is modified to accommodate magnitude-dependent dust following the prescription given by interpolating the values in Table~\ref{tab:dust_results}; the resulting adjusted LF is shown as points in Fig.~\ref{LF-MF.fig}. The Schechter function (Schechter 1976) fit to these data is shown with the thick solid line;  the Schechter function parameters are \Mstar = $-21.0^{+0.5}_{-0.6}$, \phistar = $(-2.74 ^{+0.28}_{-0.35}) \times 10^{-3}$ Mpc$^{-3}$, and $\alpha$ = $-1.47^{+0.24}_{-0.21}$.  Using the observed distribution of \ebv\ values (as did Reddy \& Steidel 2009) rather than medians (as was done above) produces almost identical results. Some differences with the original \zs2.3 KDF LF are apparent. 

Figure~\ref{LF-MF.fig} compares the updated, variable dust LF with the original constant-dust KDF fiducial result shown as the continuous light-blue line.  The most dramatic change from the original result lies in the slope of the faint end of the LF:   there are more faint BX galaxies than in the original LF, and the faint-end slope of the Schechter function has steepened from the $\alpha$=$-1.20^{+0.24}_{-0.22}$ to  $\alpha$ = $-1.47^{+0.24}_{-0.21}$.  These changes are to be expected given that for low dust values the original KDF work predicted smaller \Veff\ end hence higher LF normalizations. Given the evidence for luminosity-dependent extinction (\S~\ref{sec:dust}), the updated \zs2.3  LF presented here should be regarded as more accurate than the original KDF one. 

Figure~\ref{LF-MF.fig} also shows the \zs2 LF determined by Gabasch et al.\ (2004) from their FORS Deep Field, and that determined by Reddy \& Steidel (2009) from their large, deep BX samples.  All the LFs agree reasonably well at the bright end, but differ at the faint end.  The faint end of original KDF LF was already steeper than the Gabasch et al.\ (2004) FORS LF, and the improved, dust-corrected KDF result is steeper still.  However, the adjusted KDF faint-end slope is not as steep as that of Reddy \& Steidel (2009), even if the more extreme models are used (gray lines) instead of the fiducial one.  At $M_{1700}=-18$ the Reddy \& Steidel (2009) LF contains 2--3 times more BX galaxies than does the adjusted KDF one.  There are several possible reasons for this difference: Although Steidel \& Reddy (2009) used the same BX sample selection technique as did the KDF, they (1) had more area, and (2) used a different LF estimator, and (3) used mag-dependent extinction corrections to model galaxy SEDs in estimating survey volumes.  Reddy \& Steidel (2009) suggest that the discrepancy is due to the difference in LF estimators; this suggestion remains to be tested by analyzing a common dataset using the two different techniques.

\begin{figure}
\includegraphics[width=8cm]{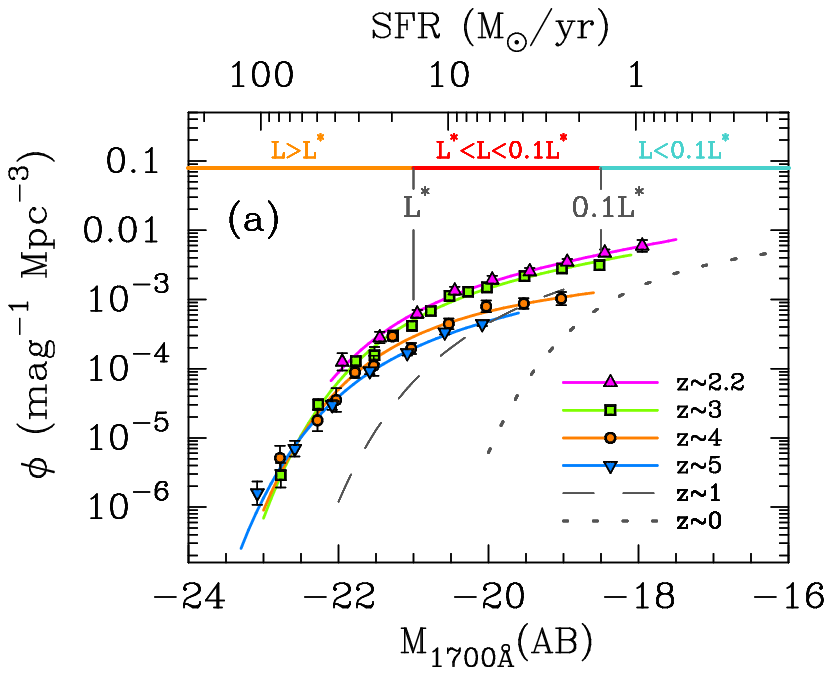}
\includegraphics[width=8.5cm]{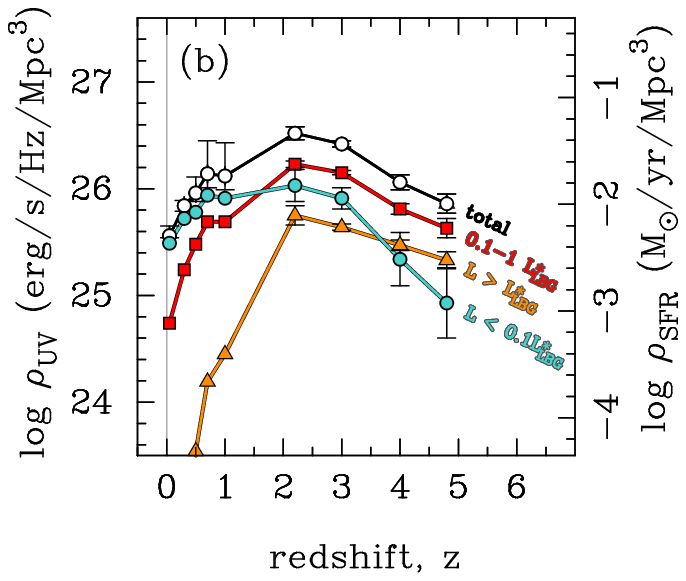}
\caption{
The luminosity function (top panel) and luminosity density (bottom panel) updated from the original Sawicki \& Thompson (2006a, 2006b) KDF analyses at \zs2.3 to account for the magnitude-dependent dust corrections on \Veff.  The \zs3 and 4 values are unchanged from the original KDF, where magnitude-dependent dust does not affect \Veff\ estimates significantly.  The \zs5 LF is from the Subaru Suprimecam study by Iwata et al.\ (2007) and the \zs1 and 0 LFs are from the GALEX results of Wyder et al. (2005) and Arnouts et al. (2005). In the bottom panel the comoving UV luminosity densities are calculated from the LFs of the top panel.  The total luminosity density is shown along with its contributions due to galaxies in different luminosity bins. 
}
\label{fig:updatedLFandLD}
\end{figure}

In the original KDF work Sawicki \& Thompson (2006a) discussed the evolution of the UV luminosity function, and Sawicki \& Thompson (2006b) examined the contributions of different luminosity galaxies to the UV luminosity density of the Universe.  Both these papers acknowledged the possibility of a dust-dependent systematic uncertainty at \zs2.3 in these measurements.  Now both the LF evolution plot and the UV luminosity density plot can be updated at \zs2.3 and these updated measurements are shown in Fig.~\ref{fig:updatedLFandLD}.  Overall, the luminosity-downsizing picture presented in the KDF papers remains unchanged: the most luminous galaxies disappear most quickly, and the least luminous ones persist the longest.  Sub-\Lstar\ galaxies (where \Lstar\ = $-$21.0 as at \zs3) dominate the luminosity density at all redshifts.  

As Fig.~\ref{fig:updatedLFandLD}(b) shows, galaxies fainter than the \zs2.3 \Lstar\ dominate the UV luminosity density of the Universe at all epochs. At the \zs2.3 epoch that is the focus of this paper, sub-\Lstar\ galaxies contribute 84\% of the rest-frame 1700\AA\ luminosity density that escapes from galaxies into intergalactic space, a result that is consistent with the earlier findings of Sawicki \& Thompson (2006b) and of Reddy \& Steidel (2009).  Assuming that absorption by neutral hydrogen gas is not stronger in sub-\Lstar\ galaxies than in more luminous objects, it is then reasonable to conclude that sub-\Lstar\ galaxies are responsible for the bulk of the ionizing photons escaping into the intergalactic medium at these epochs.  Sub-\Lstar\ galaxies thus may make a key contribution to keeping the Universe ionized at \zs2.

\subsection{Stellar mass function and the stellar mass density of the Universe}\label{sec:SMF}

\begin{figure}
\includegraphics[width=8cm]{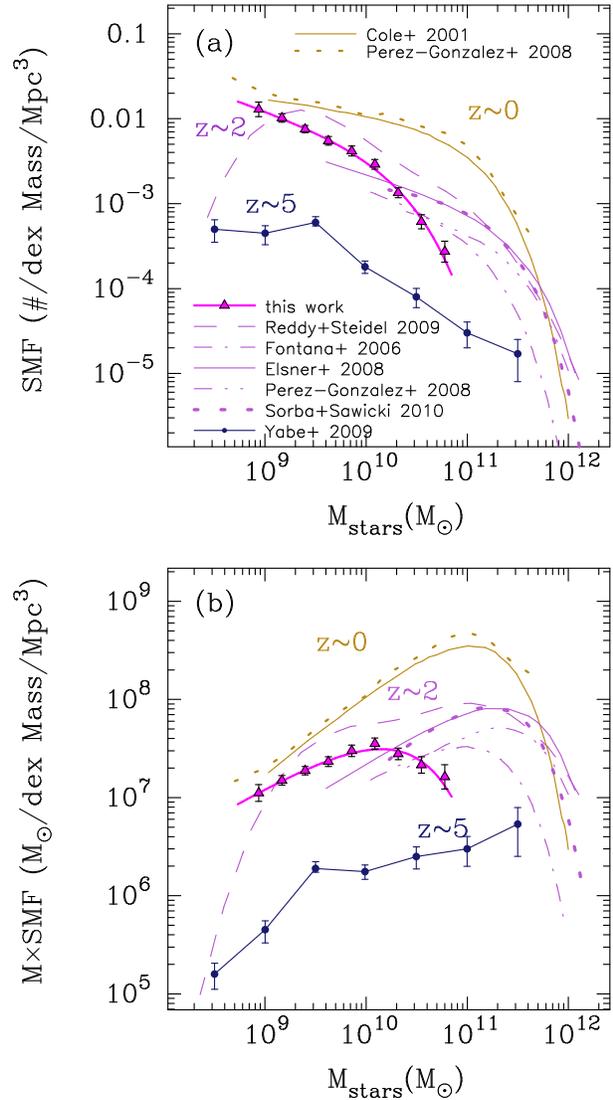}
\caption{
The stellar mass function. The top panel shows measurements of the stellar mass function at different redshifts.  At \zs2 several determinations are shown, including that by Reddy \& Steidel (2009) which, at the low-mass end, uses an earlier version of the HDF-BX UV luminosity-to-stellar mass conversion presented in the present work (Sawicki et al.\ 2007). Where needed, all the results have been converted to the Salpeter (1955) IMF. The bottom panel shows the mass-weighted stellar mass functions. A clear picture of the build-up of stellar mass over cosmic time can be seen. 
}
\label{fig:MF} 
\end{figure}

The correlation between UV luminosity and stellar mass seen in Fig.~\ref{Mstars-vs-mag.fig} (and Eq.~\ref{eq:mass-mag}) invites a straightforward conversion from the UV luminosity function to a stellar mass function.  This conversion can be accomplished by using Eq.~\ref{eq:sfr-mass} to recalibrate the axes on a UV LF plot, as is done in Fig.~\ref{LF-MF.fig} where the top and right axes recast the observations in terms of a stellar mass function.  After this conversion, it is clear that at \zs2 the number of low-mass BX galaxies is significantly higher than that of high-mass ones: by number, low-mass galaxies are a very important population at \zs2. 

The MF thus obtained is also plotted with symbols and thick solid line in Fig.~\ref{fig:MF}(a). Several other \zs2.2 stellar mass functions are also shown, namely those of Reddy \& Steidel (2009; \zs2.3), Fontana et al.\ (2006; \zs2.5), Elsner et al.\ (2008; \zs2.0), P\'erez-Gonz\'alez et al.\ (2008; \zs2.25), and Sorba \& Sawicki (2010; \zs2). With the exception of the Reddy \& Steidel (2009) mass function, all these mass functions are based on mass-selected samples; the Reddy \& Steidel (2009) mass function uses BX selection, with direct stellar mass measurements from SED fitting of bright galaxies and an earlier version of the  HDF-BX SFR-to-mass conversion (Sawicki et al.\ 2007) to estimate the contribution of faint galaxies.  Consequently, their results are not entirely independent of the present ones, but share a degree of commonality at the low-mass end. Crucially for probing low-mass galaxies, the present HDF work (and the related results of Reddy \& Steidel 2009), reach an order of magnitude in mass deeper into the low-mass end of the \zs2 population than do the other studies.  This is important given the large number of low-mass galaxies present at \zs 2. It is interesting to note that the deeper studies (present work; Reddy \& Steidel 2009) yield a larger number density of low-mass galaxies then would be inferred from extrapolations of the shallower surveys. 

At the low-mass end of the MF differences between the present MF result and that of Reddy \& Steidel (2009) are due to a combination of differences in the faint end of the input LFs (as shown in Fig.~\ref{LF-MF.fig}) and differences in the luminosity-to-mass conversion procedures.  For the latter, Reddy \& Steidel (2009) used an earlier version of Eq.~\ref{eq:sfr-mass} that was presented in Sawicki et al.\ (2007) along with their own model of magnitude-dependent dust. Consequently, their results include the potentially non-cancelling effect of two different dust corrections: one, from the SED fitting of Sawicki et al.\ 2007, to get from observed $M_{UV}$ to SFR, the other, based purely on UV slope, to get back from from SFR to $M_{UV}$).  In contrast, the procedure used in the present work is more direct as it bypasses any assumptions about dust corrections on UV luminosity by simply using the empirical relationship between directly observed UV luminosity and stellar mass (Eq.~\ref{eq:mass-mag}). For more massive galaxies, the difference in the stellar mass functions arises because Reddy \& Steidel used direct SED fits that their large sample of $\sim$ \Lstar\  BX galaxies afforded them --- an approach that is preferable to using a luminosity-mass conversion which necessarily breaks down for these more dusty, luminous objects.  Overall, however, the two MF determinations are in reasonable agreement and both emphasize that there is a large number of low-mass galaxies at \zs2.  

At the high-mass end the differences between the various \zs2 studies are quite pronounced, with differences of an order of magnitude or more in the number density of $10^{12}$\Msun\ galaxies between the various surveys.  However, regardless of which of the high-mass end results prove closer to the truth, it is clear that both high-mass and low-mass galaxies contribute significantly to the stellar mass density of the universe at \zs2. This fact is illustrated further in Fig.~\ref{fig:MF}(b) which shows the mass-weighted stellar mass function. 

Figure~\ref{fig:MF} also shows examples of stellar mass functions in the local universe (Cole et al.\ 2001; and P\'erez-Gonz\'alez et al.\ 2008). It has been known that the high-mass end of the stellar mass function was largely in place by \zs2 (e.g., P\'erez-Gonz\'alez et al.\ 2008).  However, the present data (and those of Reddy \& Steidel 2009) on the low-mass end suggest that the number density of low-mass galaxies at \zs2 was also quite high, within a factor of $\sim$2 or so of the local number density for \Mstars$\sim$10$^9$\Msun\ galaxies. If true, it would appear that the bulk of the growth since \zs2 has been in the number of mid-weight galaxies (\Mstars\ $\sim$ 10$^{11}$\Msun). 

\begin{figure}
\includegraphics[width=8cm]{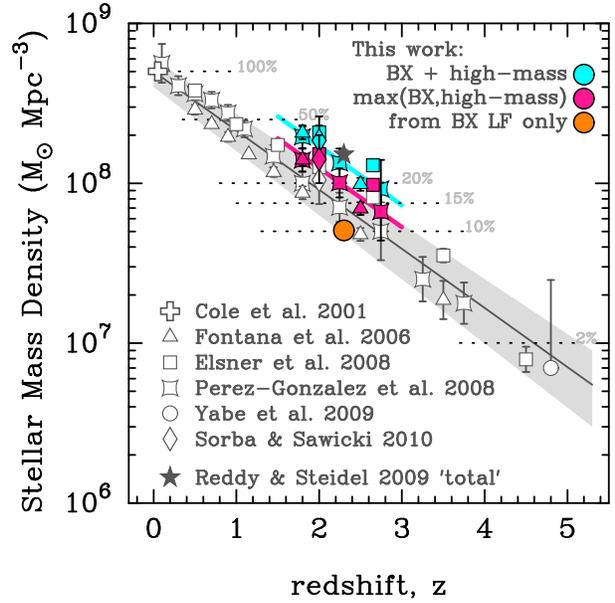}
\caption{
Evolution of the comoving stellar mass density. All measurements are adjusted to the 0.1$>$$M/M_\odot$$>$100 Salpeter IMF and integrated down to $1\times 10^8$\Msun.  The orange "from BX LF only" point is derived by converting the UV LF to a MF using the light-to-mass conversion of Eq.~\ref{eq:mass-mag}. The blue and magenta points result from combining the shallower measurements with the deeper results of the present study.  See text for more details (\S~\ref{sec:SMF}).   The horizontal dashed lines indicate the fraction of the assembled present-day stellar mass, as labelled. The gray region and line are an approximate representation of the growth of the $M_{stars} > 10^8 M_\odot$ stellar mass density as inferred by the shallower surveys (unfilled symbols).
}
\label{fig:SMD} 
\end{figure}

Figure~\ref{fig:MF} also shows the \zs5 stellar mass function of Yabe et al.\ (2009). Overall, there is clear growth in the stellar mass function across the redshift range covered, which is also reflected in the growth of the total mass density of the universe - a quantity that  has been extensively studied over the past few years. Figure~\ref{fig:SMD} presents a compilation of SMD results from several studies.  In all the data shown, the stellar mass density has been integrated down to $1\times 10^8$\Msun, although in all cases this involved an extrapolation from higher-mass data since no study at high redshift yet reaches such low-mass galaxies directly.  Integrating further, to $M=0$, would not add appreciably to the stellar mass densities because the contribution of galaxies with $M<10^8$\Msun\ is small, as can be seen in Fig.~\ref{fig:MF}(b). The open symbols at high redshift show the results of various shallower studies, while the orange point shows the mass density in low-mass galaxies (integrated down to $10^8$\Msun) as inferred from the present work.

As Fig.~\ref{fig:SMD} illustrates (open symbols), the accepted view is that stellar mass density builds up over cosmic time with perhaps $\sim$1\% of the present stellar mass density being in place at \zs5, $\sim$15\% by \zs2.3, and $\sim$50\% by \zs1. The present BX-based measurement matches that picture (orange point in Fig.~\ref{fig:SMD}).  However, neither the shallower studies, nor the present, deeper, HDF study account fully for the total mass, even with extrapolations to low or high masses. 

To account for both low-mass and high-mass galaxies properly one should combine the results of the two types of studies. The blue and magenta points show the consequence of doing this combination in two different ways.  The magenta points were obtained by taking the maximum of the low-mass\footnote{Before combining it with each of the high-mass results at $2.6 < z < 3$, the \zs2.3 low-mass stellar mass function of the present study was shifted in number density following the solid black line in Fig.~\ref{fig:MF} to reflect its likely evolution with redshift.} and high-mass stellar mass functions in each mass bin and then integrating to $M=10^8$\Msun; they likely underestimate the mass density at intermediate masses ($\sim$$3\times 10^{10}$\Msun) where both the high-mass and low-mass measurements are likely deficient. The magenta points are thus a conservative estimates of the total stellar mass densities.  The blue points were obtained by summing, in each mass bin, the low-mass and high-mass measurements and then integrating over the mass range down to $10^8$\Msun; because summing the mass functions effectively double-counts galaxies where the two mass functions are approximately equal ($\sim$$3\times 10^{10}$\Msun), the blue points are a plausible upper limit to the stellar mass density. 

The resulting "total" stellar mass densities (blue and magenta points in Fig.~\ref{fig:SMD} are higher than the extrapolations of either the shallower survey results or the present results alone.  Specifically, after correcting for the steep low-mass end of the stellar mass function, the total stellar mass densities are $\sim$1.5--2 times higher than those obtained from extrapolations of the shallower surveys. These higher stellar mass densities are in agreement with the "total" stellar mass density found by Reddy \& Steidel (2009; star symbol) using their directly-determined stellar masses at the bright end in combination with a LF-derived contribution for the low-mass end, which they obtained using their UV LF and UV-to-SFR conversion, and the SFR-to-mass conversion of Sawicki et al.\ (2007).  

Altogether, it thus seems that while the broad picture of a gradual build-up of stellar over time remains correct, the actual stellar mass density at \zs2 is higher, by a factors of $\sim$1.5--2, than those extrapolated from observations of the more massive galaxies alone. Apparently, by \zs2.3 the universe had already made $\sim$25\% of its present-day stellar mass instead of the $\sim$15\% that most of the shallower measurements suggest. In other words, the build-up of stars in the high-redshift universe has proceeded more quickly than was previously thought.

\subsection{Dust content as a function of galaxy stellar mass}\label{sec:dust-mass}

\begin{figure}
\includegraphics[width=8cm]{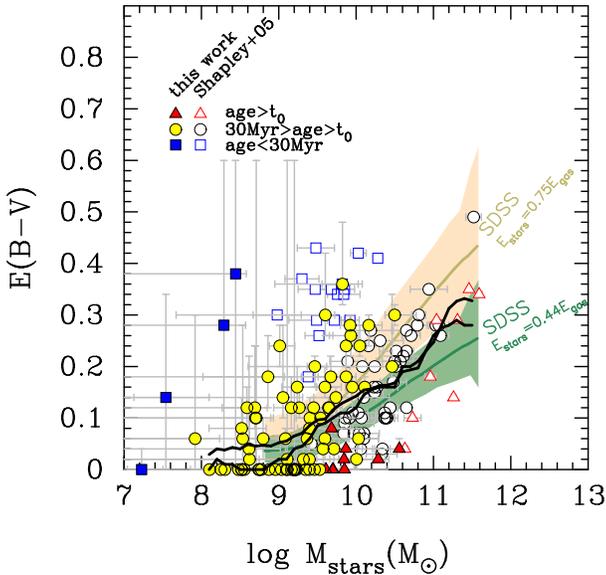}
\caption{
The dependence of color excess \ebv\ on stellar mass. Shapley et al.\ (2005) do not provide uncertainty estimates on their \ebv\ measurements, so only stellar mass uncertainties are plotted for their sample. The black curves shows the running mean and median \ebv\ values (upper and lower curves, respectively) taken with a box of width log~\Mstars=1, after excluding galaxies with age$>$$t_0$ and age$<$30 Myr.  The colored solid lines and shaded regions show two implementations of the local reddening-stellar mass correlation found in SDSS galaxies by Garn \& Best (2010):  green marks the fiducial relation while tan shows what might happen if, for example, star formation was more globally distributed within high-$z$ galaxies than is the case in local starbursts. 
}
\label{fig:ebv-vs-Mstars} 
\end{figure}

The dependence of extinction (parametrized as color excess, \ebv) on galaxy stellar mass is shown in Fig.~\ref{fig:ebv-vs-Mstars}. Overall, it appears that galaxies with higher stellar masses are also more heavily extincted. A similar trend is seen in the local universe, where several potential explanations for this phenomenon are usually offered: (1) more massive galaxies may have built up larger dust reservoirs than have the less massive ones,  (2) more actively star-forming galaxies may have larger and more dusty star-forming regions, and (3) more metal-rich galaxies may have higher dust-to-gas ratios and, consequently, higher extinctions.  Based on a large SDSS sample of local galaxies, Garn \& Best (2010) argue that stellar mass is the underlying cause of the variation in extinction between galaxies, with the SFR and metallicity correlations   resulting as a byproduct of the underlying dust-stellar mass correlation. It is plausible that this is also the case at \zs2, although the data at high redshift are not yet adequate to perform an analysis similar to that of Garn \& Best (2010) beyond the local universe. 

It is interesting to compare the \zs2 results of the present study with those for the local universe (Garn \& Best 2010).  To do so, it is necessary to account for the methodical differences between the studies, since in the present \zs2 SED-fitting study extinction results are based on stellar continua, while Garn \& Best (2010) used the Balmer decrement between nebular emission lines.  This methodological difference is significant because nebular line measurements probe sightlines to the excited gas that surrounds short-lived, intensely-radiating massive stars -- i.e., to environments that are hostile to dust --- while continuum measurements probe those to regions dominated by older, lower-mass stars that are more benign to dust. Calzetti et al.\ (2000) found that $E(B-V)_{stars} = 0.44 E(B-V)_{gas}$ in local starburst galaxies, likely reflecting such geometrical differences. Using this relation along with the Calzetti et al.\ (2000) starburst extinction curve, and adjusting for differences between IMFs in the present (Salpeter IMF) and the Garn \& Best (2010; Kroupa IMF) studies, it is then possible to plot the Garn \& Best (2010) SDSS results in Fig.~\ref{fig:ebv-vs-Mstars}. For the canonical $E(B-V)_{stars} = 0.44 E(B-V)_{gas}$ of Calzetti et al.\ (2000), these are shown using the green line and shaded region.  In the mean there is reasonable agreement between the \zs2 population and the local result. 

While in the mean the local and \zs2 relations are in reasonable agreement, the large amount of scatter seen at \zs2 implies that there are many high-$z$ galaxies that are significantly above the local relation, as well as others that are below it.  Garn \& Best (2010) suggested that the local extinction-stellar mass relation can be used to predict the extinctions of high-$z$ galaxies of known stellar mass, but doing so can be problematic for individual objects given the large scatter seen in the \ebv-\Mstars\ plot at \zs2. 

Nevertheless, the fact remains that at \zs2 there appears to exist a correlation between dust and galaxy stellar mass. It is then interesting to consider what may be its root cause.  One possible explanation may lie in the correlation between stellar mass and metallicity (Pettini et al.\ 2001; Erb et al.\ 2006).  These authors found that more massive high-z galaxies are also more metal-rich, and one could speculate that the more massive objects contain more dust because they contain larger fractions of the elements required to produce dust grains. However, this is not the only possible explanation: for example, it is also possible that lower-mass galaxies are also physically smaller and thus --- for a fixed physical gas and dust density --- they present lower column densities of both gas and dust. A similar trend would also result if galaxy sizes were constant as a function of stellar mass, but with more massive galaxies having higher densities of both stars and dust-bearing gas.  An examination of physical sizes of \zs2 galaxies as a function of their stellar mass may shed light on this issue, although such analysis is beyond the scope of this paper. Finally, it is worth pointing out again that in the local universe Garn \& Best (2010) found that the primary driver of extinction is not metallicity but rather galaxy stellar mass, with metallicity (and SFR) being indirectly linked to extinction by virtue of all of them being correlated with the more fundamental quantity that is galaxy stellar mass. Although \zs2 samples are not yet suitable for a similarly detailed analysis, it is reasonable to suppose that at high redshift it is also stellar mass, rather than metallicity, that governs the amount of extinction.

\subsection{A simple model for the growth of galaxy masses}\label{sec:sfr-mass-model}

The nearly linear relationship between ongoing star formation rate and the accrued stellar mass that is evident in Fig.~\ref{SFR-vs-Mstars.fig} (also Eq.~\ref{eq:sfr-mass}) may, at first glance, appear puzzling.  Simply put, if a galaxy is forming stars at a given rate, should it not, in short order, grow its stellar population so that it no longer lies on the observed SFR-\Mstars\ relation?  How does such a growing galaxy ``know" how to adjust its SFR as its stellar mass increases with time?  This section develops a simple model that can account for this phenomenon. 

\subsubsection{How do BX galaxies chose their SFRs?}

Let's assume that the star formation rate in a BX galaxy is simply related to the rate at which it is accreting new gas, modulated by the instantaneous, or nearly-instantaneous, star formation efficiency, \sfe. The \sfe\ here is more strictly defined as the fraction of the accreting gas that gets converted into stars, so that 
\begin{equation}\label{sfe.eq}
SFR = \dot{M}_{stars} =  f_* \dot{M}_{gas}.
\end{equation}
This assumption is similar to that made by Tilvi et al. (2009) for Lyman-$\alpha$ emitters (LAEs), but while Tilvi et al.\ included an on/off duty cycle of star formation in their LAEs, the correlation between SFR and stellar mass in BX galaxies (Fig.~\ref{SFR-vs-Mstars.fig} and \S~\ref{sec:ages}) suggests that steady-state star formation is appropriate here, making such a duty cycle unnecessary.  Note that in this model the fraction of the accreting gas that fails to be converted into stars right away, namely ($1$$-$\sfe),  does not participate in star formation but remains in gaseous form indefinitely and even possibly leaves the DM halo: it can be though of as having been shock-heated, ejected, or made otherwise unavailable for star formation on timescales of relevance.

If gas and dark matter accrete out of the background onto a galaxy together, and some fraction of the gas gets converted into stars on a reasonably short timescale, as explained above, then 
\begin{equation}\label{sfe2.eq}
SFR =  f_* (\Omega_b/\Omega_{DM})\dot{M}_{DM},
\end{equation}
 where $\dot{M}_{DM}$ is the host halo's dark matter accretion rate and $\Omega_b/\Omega_{DM}$ is the cosmic baryon-to-dark matter ratio.  Note that $f_*$ and   $\Omega_b/\Omega_{DM}$ in the above equations are degenerate and their values may vary somewhat from galaxy to galaxy, possibly in a systematic way, so long as their product remains fixed.  Indeed, in practice $f_*$ is unlikely to be constant with galaxy mass due to various physical feedback effects at play.
 
With the help of Eq.~\ref{sfe2.eq}, Eq.~\ref{eq:sfr-mass} can then be recast as
\begin{equation}\label{eq:DMrate-Mstars}
\log \left( \frac{\dot{M}_{DM}}{M_\odot yr^{-1}} \right) = 0.89 \log \left( \frac{M_{stars}}{M_\odot} \right) - 7.69 - \log \left(\frac{f_* \Omega_b}{\Omega_{DM}} \right),
\end{equation}
which implies that the stellar mass of a BX galaxy is nearly proportional to the dark matter accretion {\emph {rate}} of its halo:  $\dot{M}_{DM} \propto (M_{stars})^{0.89}$.  Note that in this model more massive halos accrete matter --- both dark and baryonic --- at a faster rate than do lower mass halos. A galaxy's SFR is proportional to the present baryonic mass accretion rate and stellar mass is proportional to the integral of the past baryonic mass accretion rate, so it follows that galaxies with high stellar masses also have high star formation rates.  This argument accounts for the observed correlation between SFR and stellar mass.

\subsubsection{Dependence on DM halo mass}

In \lcdm\ cosmologies dark matter halos accrete matter more quickly than do low mass ones. One can then ask how will a galaxy's properties depend on its DM halo mass in the model above.  Using \lcdm\ dark matter simulations, Tivli et al.\ (2009; see their Fig.~3) find that the growth rate of halos can be approximated as  
\begin{equation}\label{eq:tavli}
\log \left(\frac{\dot{M}_{DM} }{ M_\odot yr^{-1}}\right) = m \log\left( \frac{M_{DM}}{M_\odot}\right) + b, 
\end{equation}  
with slope $m$ and intercept $b$ that are nearly invariant over the redshift range $z$=6.6 -- 3.1 discussed in their work. 
Their $z=3.1$ values (the lowest redshift they show) are $m=0.85$ and $b = -6.35$  and $m$=0.8--0.9 at higher redshifts. Although they do not show the accretion relation at lower redshifts, given its invariance at higher redshifts, it seems reasonable to adopt their higher-redshift slope values, $m$=0.8--0.9, for $z$$\sim$2.2 as well. 
The similarity in slopes with Eq.~\ref{eq:DMrate-Mstars} then suggests that the observed SFR-$M_{stars}$ correlation may be simply related to the halo mass accretion rate.
Following this line of reasoning and combining equations \ref{eq:DMrate-Mstars} and \ref{eq:tavli}, while taking $m$=0.9 and $b = -6.4$  for simplicity and consistency with the lowest-$z$ simulation results, yields
\begin{equation}\label{eq:massratio-to-sfe}
0.9 \log \left(\frac{M_{stars}}{M_{DM}} \right)= \log (f_*) + \log \left(\frac {\Omega_b }{ \Omega_{DM}} \right)+1.3
\end{equation}
This equation simply states that the stellar-to-dark matter mass ratio in galaxies is related to the cosmic baryon-to-dark matter ratio via the instantaneous efficiency at which gas is converted into stars (\sfe). The equation also provides an empirical calibration of this conversion rate, \sfe\ determined from the BX stellar mass and SFR measurements combined with theoretical dark halo growth rates.  Moreover, 
the fact that the slopes of the empirical relation for stellar mass growth (Eq.~\ref{eq:DMrate-Mstars}) and the $\Lambda$CDM model-derived one for halo growth (Eq.~\ref{eq:massratio-to-sfe}) are similar suggests that the star formation efficiency may be constant over the mass range considered here. 

\subsubsection{Star formation efficiency in BX galaxies}

With Eq.~\ref{eq:massratio-to-sfe} in hand and a knowledge of stellar-to-dark matter ratio in \zs2 galaxies, it should then be possible to estimate the star formation efficiency.  Deriving the $M_{stars}/M_{DM}$ ratio requires a combination of stellar mass and clustering measurements for \zs2 BX galaxies, and such measurements have not yet been done concurrently.  However, the ratio can be estimated by combining separate measurements of luminosity dependence of stellar mass (such as in the present work) with measurements of the luminosity dependence of halo mass (e.g, Savoy et al.\ 2011).  Simply put, one ties the two measurements, performed on similar but separate samples, by means of a common UV luminosity scale.  The situation is complicated by the fact that while at higher redshifts, $z\ga3$, there appears to exist a well-established correlation between UV luminosity and halo mass (e.g., Giavalisco \& Dickinson 1998, Ouchi et al.\ 2001, Adelberger et al.\ 2005), the results for UV-selected galaxies at \zs2 are still sparse and contradictory: Adelberger et al.\ (2005) find a correlation between UV luminosity and clustering strength among BX galaxies, while Savoy et al.\ (2011), using an identically-selected deeper sample of BX galaxies in the Keck Deep Fields (Sawicki \& Thompson 2005), find a more complex relation that they interpret as due to confusion of the clustering signal by very massive halos ($M_{DM}>10^{12}M_\odot$) whose resident galaxies are in the process of shutting down their star formation (i.e., a manifestation of downsizing).

Despite this confused \zs2 clustering picture, it is nevertheless possible to put constraints on the $M_{stars}/M_{DM}$ ratio and hence on the star formation efficiency.  Combining the stellar masses from the present study with the deep BX clustering measurements from Savoy et al.\ (2011) give $M_{stars}/M_{DM} \sim $0.01--0.005 (see Savoy et al.\ 2011 for more details). Combining them instead with the clustering results of Adelberger et al.\ (2005) results in $M_{stars}/M_{DM} \sim 0.06$.  With these two results in hand, and assuming the cosmic baryon-to-dark matter mass ratio of $\Omega_b/\Omega_{DM} = 0.2$ (e.g., Komatsu et al.\ 2009), Eq.~\ref{eq:massratio-to-sfe} then tells us that the star formation efficiency of \zs2 BX galaxies is \sfe$\sim$0.2--2\%, which is similar to the $\sim$2.5\% derived for Ly$\alpha$ emitters by Tilvi et al.\ (2009). Strictly speaking, because $f_*$ and $\Omega_b/\Omega_{DM}$ are degenerate and their values may vary for individual galaxies, the value of $f_*$ derived above is not necessarily expected to be constant from galaxy to galaxy.  Nevertheless, the model suggests that only a small fraction of the gas that falls into a BX galaxy's halo ends up condensing into stars, with the bulk of the accreted gas remaining unconsumed.

 \subsubsection{Limitations and implications for star formation histories}\label{sec:limitations}

The simple model described above is remarkably successful at accounting for the observed correlation between star formation rate and accrued stellar mass in \zs2 UV-selected star-forming galaxies. The model would, clearly, fail were it to be applied to ``red and dead" galaxies at these redshifts such as pBzK galaxies (Daddi et al.\  2004) or DRGs (Franx et al.\ 2003). In such galaxies star formation has largely shut down already and even though material is likely still being accreted onto them, some mechanism prevents it from being formed into stars. In the simple picture suggested in the model this would be equivalent to a star formation efficiency that is much lower than that in the present sample.  

The simple model likely also fails in UV-selected galaxies at higher redshifts, \zs5. While a SFR-stellar mass relation is observed there (Yabe et al.\ 2009), the assumption of constant SFR --- as required by the dependence of SFR on halo mass accretion rate --- would overproduce the masses observed in \zs2 star-forming galaxies of similar SFR (Yabe et al.\ 2010).  This overshoot could potentially be resolved by invoking star formation rates that decline with time, but this approach would require either an equally declining mass accretion rate (an unlikely proposition), or a time-variable star formation efficiency.  Attractively, the \zs5$\rightarrow$2 mass overshoot could possibly also be resolved by an {\it increasing} SFR rate that would make galaxies more massive {\it and} more star-forming over time and would be naturally consistent with the idea of increasing mass accretion rates that follows from the growing halo paradigm.  Such increasing star formation histories have recently been discussed (e.g.,  Lee et al.\ 2010,  2011; Maraston et al.\ 2010;  Papovich et al.\ 2010) and thus appear attractive on other grounds as well.  In this respect, low-mass galaxies may be able to provide a clearer picture than that afforded by their more massive counterparts which likely suffer more strongly from a variety of effects that act to complicate their star formation histories thereby making the link to hierarchical models less direct.

In conclusion, despite its limitations, 
the simple model is nevertheless remarkable in that it successfully explains the observed correlation between SFR and accrued stellar mass in an important class of galaxies at \zs2 and it does so by invoking a simple and quite plausible mechanism whereby star formation rate is related directly to the growth rate of dark matter halos that is a fundamental feature of \lcdm\ cosmologies.

\section{SUMMARY AND CONCLUSIONS}\label{sec:conclusions}

This paper presents the results of an SED-fitting study of the stellar populations of sub-\Lstar\ UV-selected galaxies at \zs2.  The main results of this study can be summarized as follows. 

\begin{enumerate}

\item {\it $L_{UV}$-\Mstars\ correlation.} Deep optical-IR observations show that there is a correlation between UV luminosity and stellar mass in sub-\Lstar\ BX galaxies (Eq.~\ref{eq:mass-mag}). The existence of this correlation suggests that star formation histories in many sub-\Lstar\ BX galaxies may be approximately constant with time, rather than dramatically variable. In addition to revealing such physical insights, the \Muv - \Mstars\ correlation allows a useful approximate conversion between rest-frame UV (observed optical) luminosities and stellar masses of BX galaxies which can be used to estimate stellar masses when infrared observations are unavailable. 

\item {\it Stellar mass function and stellar mass density of the Universe.}  The low-mass end of the stellar mass function at \zs2.3 is estimated from the UV LF function by means of the empirical \Muv\ - \Mstars\ relation. It is found that there are many low-mass galaxies at this epoch ---the low-mass end of the BX SMF is steeper than predicted by extrapolations of shallower surveys at \zs2, with more low-mass galaxies present at this epoch than such extrapolations predict. These numerous low-mass galaxies contribute significantly to the stellar mass density at \zs2.3, increasing its values by a factor of $\sim$1.5 compared to extrapolations from higher-mass data alone.  This upward revision means that by \zs2.3 the Universe has already made $\sim$22\% of its present-day stellar mass, as compared to only $\sim$15\% that the extrapolations of the shallower surveys would suggest. Apparently the build-up of stellar mass in the universe has proceeded somewhat more quickly than previously thought. 

\item {\it Correlation between UV luminosity and interstellar extinction.}  Sub-\Lstar\ BX galaxies appear to have less dust than their $\sim$\Lstar\ counterparts. For example, more than half the UV photons produced by a typical $M^*$+3 (\Mstars$\sim$$10^9$\Msun) galaxy escape into intergalactic space, in contrast to $L$$\sim$\Lstar\ BX galaxies from which only about 1 in 5 UV photons escape absorption by dust. Sub-\Lstar\ BX galaxies are thus very important contributors to the UV luminosity density budget at \zs2. 

\item {\it Correlation between interstellar extinction and stellar mass.} The data show that more massive galaxies \zs2 galaxies tend to be more dusty than lower-mass galaxies.  This trend is similar to that found in the local universe, albeit with significantly larger scatter. 

\item {\it Updated UV luminosity function and luminosity density of the Universe.} The dependence of reddening on UV luminosity results in a revision of the effective survey volume, \Veff, of the KDF survey at \zs2.3 (Sawicki \& Thompson 2006a) and hence a revised \zs2.3 UV luminosity function.  The updated LF has a steeper faint-end slope, with $\alpha=-1.47^{+0.24}_{-0.21}$ (compared to the previous $\alpha=-1.20^{+0.24}_{-0.22}$), although this revised slope is not as steep as the $\alpha=-1.73\pm0.13$ reported by Reddy \& Steidel (2009).  Nevertheless, it is clear that sub-\Lstar\ galaxies at \zs2 are very numerous and contribute significantly to the UV luminosity density at this epoch, producing $\sim$84\% of the emergent 1700\AA\ luminosity density.  If absorption by neutral hydrogen in sub-\Lstar\ galaxies is not stronger than in brighter objects then sub-\Lstar\ galaxies make a key contribution to keeping the universe ionized at these redshifts.

\item {\it SFR-\Mstars\ correlaton.} A correlation between star formation rate and stellar mass exists in \zs2.3 galaxies, and this correlation extends over three orders of magnitude in stellar mass. This nearly 1-to-1 (in log-log space) correlation suggests that the rate at which a sub-\Lstar\ galaxy form new stars is related to its past star formation history, and --- more precisely --- to the amount of stars that the galaxy has already formed. 

\item {\it A scenario for the growth of galaxy stellar masses.} The existence of a SFR-\Mstars\ relation over a wide range of stellar masses among \zs2 star-forming galaxies poses the challenge:  how to star forming galaxies ``know" that they should have SFR $\propto$ $\int SFR(t)$~$dt$?  A simple model successfully explains this phenomenon:  the model simply assumes that SFRs are proportional to baryon accretion rates, which in turn are proportional to dark matter accretion rates.  Since in the $\Lambda$CDM paradigm DM accretion rates scale with halo mass, it naturally follows that both SFRs and stellar masses will correlate via the masses of the underlying DM halos.  Based on the scaling of halo accretion rates with mass, the model successfully accounts for the slope of the observed BX galaxy  SFR-\Mstars\ correlation, and finds that the instantaneous star formation efficiency (i.e., the fraction of accreting baryons that condenses into stars) is only about one percent or so. The key ingredient of this model --- that SFRs are governed by the (continually growing) DM masses of galaxies, has the potential to naturally account for the increasing star formation histories that have recently been suggested for high-$z$ galaxies. 

\end{enumerate}

Sub-\Lstar\ galaxies at high redshift account for significant amounts of star formation in the Universe and contain significant amounts of its stellar mass.  At the same time, they are not simply scaled-down copies of their more luminous \Lstar\ cousins, as is evidenced, for example, by the presence of differential, luminosity-dependent extinction. While high-$z$ sub-\Lstar\ galaxies are challenging to study because of their faintness, understanding their physical nature holds the potential to understand a different physical regime in galaxy formation than that probed by \Lstar\ and super-\Lstar\ objects. Moreover, as was illustrated in this paper, the ability to span several orders of magnitude in stellar mass gives the ability to tease out important hidden trends; it also enables differential, comparative measurements of galaxy properties, whose results are more trustworthy than those of absolute measurements. Sub-\Lstar\ galaxies are thus both important and useful denizens of the Universe, and we need to continue to strive to better understand what they tell us about the history of structure formation in the Universe.

\section*{Acknowledgments}

This work was supported financially by the Natural Sciences and Engineering Research Council (NSERC) of Canada, the Canadian Space Agency, and by NASA.  Thanks are due to Kouji Ohta and Daniela Calzetti for useful discussions, Jerzy Sawicki for a careful reading of the manuscript,  and the anonymous referee for several useful suggestions. Parts of the analysis presented here made use of the Perl Data Language (PDL) that has been developed by K.\ Glazebrook, J.\ Brinchmann, J.\ Cerney, C.\ DeForest, D.\ Hunt, T.\ Jenness, T.\ Luka, R.\ Schwebel, and C. Soeller. PDL provides a high-level numerical functionality for the perl scripting language (Glazebrook \& Economou, 1997) and can be obtained from http://pdl.perl.org.  During this work occasional use was also made of the convenient on-line cosmology calculator developed by Wright (2006).

\bibliographystyle{mn2e}

\bsp

\label{lastpage}

\end{document}